\documentclass[submission,copyright,creativecommons]{eptcs}
\providecommand{\publicationstatus}{To appear in EPTCS.}

\usepackage[leqno,fleqn,intlimits]{amsmath}
\usepackage{amsfonts}
\usepackage{amssymb}
\usepackage{amstext}
\usepackage{epsfig}
\usepackage{graphicx}
\usepackage[all,2cell,knot,poly]{xy}

\newcommand{\zb}{\hspace{-0.5pt}}

\newcommand{\True}{T}
\newcommand{\False}{F}

\newcommand{\Nil}{\texttt{[]}}

\newcommand{\append}{~\texttt{++}~}
\newcommand{\Prog}{\pi_0}
\newcommand{\Syn}{\pi_s}

\newcommand{\Fn}[1]{{\mathbb F}_{#1}}

\newcommand{\precceq}{\preccurlyeq}

\newcommand{\proptoeq}{\;\underline{\propto}\;}
\newcommand{\notpropto}{\rlap{~{$\not$}}\proptoeq}

\newcommand{\tur}{\;\triangleleft\;}
\newcommand{\turprec}{\triangleleft\;\circ\precceq}

\newcommand{\nottur}{\;\ntriangleleft\;}

\newcommand{\expr}[1]{#1{}}
\newcommand{\var}[2]{\mathit{#1}#2}
\newcommand{\patts}[2]{#1{,~#2}}
\newcommand{\threepatts}[3]{#1{,~#2}{,~#3}}

\newcommand{\patt}[3]{#1{~#2{#3}}}
\newcommand{\app}[3]{#1{~#2{#3}}}
\newcommand{\fapp}[3]{\var{#1}(~#2\;)#3}

\newcommand{\args}[2]{#1{,~#2}}
\newcommand{\threeargs}[3]{#1{,~#2}{,~#3}}

\newcommand{\brackets}[2]{(#1{)#2}}

\newcommand{\fsent}[3]{\begin{array}[t]{@{\hspace*{1mm}}l@{\hspace*{1mm}}c@{\hspace*{1mm}}l@{\hspace*{1mm}}}{\var{#1}{(~}} #2{~)} & \Rightarrow & #3{;}{}\end{array}}
\newcommand{\lfsent}[3]{\begin{array}[t]{@{\hspace*{1mm}}l@{\hspace*{1mm}}c@{\hspace*{1mm}}l@{\hspace*{1mm}}}{\var{#1}{(~}} #2{~)} & &\\ {\hspace*{35pt}} \Rightarrow #3{;}{} & &\end{array}}
\newcommand{\fnew}{\begin{array}[t]{@{\hspace*{1mm}}l@{\hspace*{1mm}}c@{\hspace*{1mm}}l@{\hspace*{1mm}}} & & \end{array}}

\newcommand{\sentshift}{\hspace{20pt}}
\newcommand{\shiftfsent}[3]{\fsent{{\sentshift}#1}{#2}{#3}}

\newcommand{\shiftlfsent}[3]{\begin{array}[t]{@{\hspace*{1mm}}l@{\hspace*{1mm}}c@{\hspace*{1mm}}l@{\hspace*{1mm}}}\sentshift{\var{#1}{(~}} #2{~)} & &\\ {\hspace*{35pt}} \sentshift\Rightarrow #3{;}{} & &\end{array}}

\newcommand{\psent}[2]{\begin{array}[t]{@{\hspace*{1mm}}l@{\hspace*{1mm}}c@{\hspace*{1mm}}l@{\hspace*{1mm}}}{{~~}} #1{~} & \rightarrow & #2{~.}{}\end{array}}

\newcommand{\pprop}[1]{\begin{array}[t]{@{\hspace*{1mm}}l@{\hspace*{1mm}}c@{\hspace*{1mm}}l@{\hspace*{1mm}}}{{~~}} #1{~}\end{array}}

\newcommand{\enapp}[3]{\mathfrak{{#1}(}~#2\;\mathfrak{)}#3}

\newtheorem{proposition}{Proposition}
\newtheorem{corollary}{Corollary}
\newtheorem{definition}{Definition}

\newtheorem{remark}{Remark}

\title{Verification of Programs via Intermediate Interpretation}

\author{Alexei P. Lisitsa
\institute{Department of Computer Science,\\ The University of Liverpool
}
\email{{a.lisitsa@liverpool.ac.uk}}
\and
Andrei P. Nemytykh
\institute{Program Systems Institute,\\ Russian Academy of Sciences\thanks{The second author was  supported by RFBR, research project No. 17-07-00285\_a, and 
Russian Academy of Sciences, research project No. 0077-2014-0030.
}
}
\email{\quad {nemytykh@math.botik.ru}}
}
\def\titlerunning{Verification of Programs via Intermediate Interpretation}
\def\authorrunning{A.\,P. Lisitsa, A.\,P. Nemytykh}

\begin{document}
\maketitle

\begin{abstract}
We explore an approach to verification of programs via program transformation applied to an interpreter of a programming language.
A specialization technique known as Turchin's supercompilation is used to specialize some interpreters with respect to the program models.
We show that several safety properties of functional programs modeling a class of cache coherence protocols can be proved by a supercompiler and compare
the results with our earlier work on direct verification via supercompilation not using intermediate interpretation.

Our approach was in part inspired by an earlier work by E.\,De~Angelis et al. (2014-2015) where verification via program transformation and intermediate interpretation was studied in the context of specialization of constraint logic programs. 


\end{abstract}

\section{Introduction}\label{sec:Introduction}
We show that a well-known program specialization technique
called the first Futamura projection \cite{Futamura:1971, Turchin:Report20, Jones:IntSpecialisation} 
can be used for indirect verification of some safety properties.
We consider 
 functional programs modeling a class of non-deterministic parameterized computing systems specified in a language 
that differs 
from the object programming language treated by the program specializer. 
Let a specializer transforming programs be written in a language ${\cal L}$ and an interpreter \texttt{Int}$_{\cal M}$ of a language ${\cal M}$, 
which is also implemented 
in ${\cal L}$, be given. Given a program \texttt{p$_0$} written in ${\cal M}$, the 
task is to specialize the interpreter \texttt{Int$_{\cal M}$(p$_0$,d)} with respect to 
its first argument, while the data \texttt{d} of the program \texttt{p$_0$} is unknown.

Our interest in this task has been inspired
by the following works \cite{FioPettProSen:Ver-IterSpec14, FioPettProSen:Ver-Linear-Fib15}. 
The authors work in terms of \emph{constraint logic programming} (CLP), where the constraint language is the linear arithmetic inequalities imposed on integer values of variables. They use partial deduction \cite{Leuschel:04} 
and CLP program specialization \cite{FioPettPro:01, VeriMAP:2014} methods for specializing an interpreter of a C-like language with respect to given programs, aiming at verification of the C-like imperative specifications with respect to the postconditions defined in CLP and defining the same functions (relations) as done by the corresponding C-like programs.
Additionally to the CLP program specialization system developed by 
E.\,De~Angelis et al.
and called VeriMAP \cite{VeriMAP:2014} they use also
external satisfiability modulo theories (SMT) solvers.
We would also refer to an earlier work by J.\,P. Gallagher et al. \cite{GH:1998} 
proposing 
a language-independent method for analyzing the imperative programs via intermediate interpretation by 
a logic programming language. 
Note that the transformation examples given in the papers \cite{GH:1998, FioPettPro:01, VeriMAP:2014} 
 presenting the above mentioned approaches
deal with neither function nor constructor application stack in the interpreted programs.

In this paper we focus our attention on 
self-sufficient methods for specialization of functional programs, aiming at proving some \emph{safety} properties of the programs. 
We consider a program specialization method called Turchin's supercompilation \cite{Tur:86, Turchin:Report20, SGJ:96, Jonsson_Nordlander} and study potential capabilities of the method for verifying the safety properties of the functional programs modeling a class of non-deterministic parameterized cache coherence protocols \cite{Delzanno-Param:00}.
We use an approach to functional modeling of non-deterministic computing systems, first presented by these authors in \cite{Lis_Nem:CSR07, Lis_Nem:Programming, Lis_Nem:IJFCS08}. The simple idea behind the approach is as follows. Given a program modeling a deterministic computing system, whose behavior depends on and is controlled by an input parameter value, let us call for an oracle producing the input value. Then the meta-system including both the program and the external oracle becomes non-deterministic one. And vice versa, given a non-deterministic system, 
one may be concerned about the behavior 
of the system only along one possible path of the system evaluation. In such a case, the path of interest may be given as an additional input argument of the system, forcing the system to follow along the path. Dealing with an unknown value of the additional parameter one can study any possible evolution of the system, 
for example, aiming at verifying some properties of the system.

Viability of such an approach to verification has been demonstrated in previous works using supercompilation as a program transformation and 
analysis technique 
\cite{Lis_Nem:CSR07, Lis_Nem:Programming, Lis_Nem:IJFCS08, Lis_Nem:Protocols07,Klimov:PetriNets12}, where it was applied to safety verification of program models of parameterized 
 protocols and Petri nets models.
Furthermore, the functional program modeling and supercompilation have been used to specify and verify cryptographic protocols, and in the case of insecure protocols a supercompiler was utilized in an interactive search for the attacks on the protocols 
 \cite{AhmedLisitsaNemytykh:NSPK,ANepeivoda:PingPong}.
In these cases the supercompiler has been used for specializing the corresponding program models, aiming at moving the safety properties of interest from the semantics level of the models to simplest syntactic properties of the residual programs produced by the supercompiler. Later this approach was extended by G.\,W. Hamilton for verifying a wider class of 
temporal properties of reactive systems \cite{Hamilton:VPT2015, Hamilton:VPT2016}.

Given a specializer transforming the program written in a language ${\cal L}$ and used for program model verification, in order to mitigate the limitation of the specification language ${\cal L}$, in this paper we study potential abilities of the corresponding specialization method for verifying the models specified in another language ${\cal M}$.
We analyze the supercompilation algorithms allowing us crucially to remove the interpretation layer and to verify indirectly the safety properties. The corresponding experiments succeeded in verifying some safety properties of the series of parameterized cache coherence protocols specified, for example, in the imperative WHILE language by N.\,D. Jones \cite{Jones:Complex}. Nevertheless, in order to demonstrate that our method is able to deal with non-imperative interpreted programs, we consider the case when a modelling language ${\cal M}$ is a non-imperative subset of the basic language ${\cal L}$. On the other hand, that allows us to simplify the presentation.
In order to prove the properties of interest, some of the program models 
used in the experiments require one additional supercompilation step (i.e., the corresponding residual programs should be supercompiled once again\footnote{Note that the method presented in the papers \cite{FioPettProSen:Ver-IterSpec14, FioPettProSen:Ver-Linear-Fib15} mentioned above sometimes requires a number of iterations of the specialization step  
and the number is unknown.
}). 

The considered  class of cache coherence protocols effectively forms a benchmark on which various methods for parameterized verification have been tried   \cite{Pong:1997, Delzanno-Param:00, Emerson2003, FisherKL05, Lis_Nem:Programming, Lis_Nem:IJFCS08, Lisitsa:ATVA10}. 
In \cite{Lis_Nem:Programming, Lis_Nem:IJFCS08} we have applied direct verification via supercompilation approach without intermediate interpretation.  
The corresponding models of these and others parameterized protocols 
may be very large and the automatic proofs of their safety properties may have very complicated structures. 
See, for example, the structure of the corresponding proof \cite{Lis_Nem:Protocols07} produced by the supercompiler SCP4 \cite{N:03, Nemytykh:SCP4book, NT:00} for the functional program model of the parameterized Two Consumers - Two Producers (2P/2C) protocol for multithreaded Java programs \cite{Begin:BABYLON}. 
Taking that into account, the experiments presented in this paper can also be considered as a partial verification of the intermediate interpreters \texttt{Int$_{\cal M}$(p,d)} used in the experiments. That is to say, a verification of the interpreters with respect to the subset of the input values of the argument \texttt{p}, being the program models of the cache coherence protocols.

The program examples given in this paper were specialized by the supercompiler SCP4 \cite{N:03, Nemytykh:SCP4book, NT:00}, which is a program specilalizer based on the supercompilation technique.
We present our interpreter examples in a variant of a pseudocode for a functional program while real supercompilation experiments with the programs were done in the strict functional programming language Refal \cite{Turchin:Refal5}\footnote{The reader is welcome to execute several sample Refal programs and even any program written by the user directly from the electronic version of the Turchin book.}, 
\cite{Refal5:PZ} being both the object and implementation language of the supercompiler SCP4.
One of advantages of using supercompilation, instead of other forms of partial evaluation or CLP specialization, is the use of Turchin's relation (Section \ref{subsec:Turchin}, see also \cite{Turchin:Generalization88, Nemytykh:SCP4book, AntoninaNepeivoda:VPT2016}) defined on function-call stacks, where the function calls are labeled by the times when they are generated by the unfold-fold loop. This relation is responsible for accurate generalization of the stack structures of the unfolded program configurations. It is \emph{based on global properties} of the path in the corresponding unfolded tree rather than on the structures of two given configurations in the path. Turchin's relation both stops the loop unfolding the tree and \emph{provides 
a guidance of how a given call-stack structure has to be generalized}. Proposition \ref{prop:Match} proven in this paper shows that a composition of the Turchin and Higman-Kruskal relations may prevent generalization of two given interpreter configurations encountered inside one big-step of the interpreter. Such a prevention from generalization is crucial for optimal specialization of any interpreter w.r.t.  a given program.

This paper assumes that the reader has basic knowledge of concepts of functional programming, pattern matching, term rewriting systems, and program specialization.

\vspace{-5pt}
\paragraph{The contributions of this paper are:}
(1)	
We have developed a method aiming at uniform reasoning on properties of configurations' sequences that are encountered in specializing an interpreter of a Turing complete language.
(2)	In particular, we have proved the following statement. Consider specialization of the interpreter with respect to any interpreted program from an infinite program set 
that is large enough to specify a series of parameterized cache coherence protocols, controlled by a composition of 
the Turchin (Section \ref{subsec:Turchin}) and Higman-Kruskal (Section \ref{subsec:Higman-Kruskal}) relations. 
Given a big-step of the interpreter to be processed by the unfold-fold loop, we assume 
that neither generalization nor folding actions were done by this loop up to the moment considered. Then any two non-transitive (Section \ref{subsec:Strategy}) big-step internal configurations $C_1, C_2$ are prevented from both generalization and folding actions.  
(3)	We have shown that supercompilation controlled by the composition of the relations above is able to verify  some safety properties of the series of parameterized cache coherence protocols via intermediate interpretation of their program models. Note that these program specifications include both the function call and constructor application stacks, where the size of the first one is uniformly bounded on 
the value of the input parameter while the second one is not. 
Unlike VeriMAP \cite{VeriMAP:2014} our indirect verification method involves no post-specialization unfold-fold.

The paper is organized as follows. In Section \ref{sec:Self-Interpreter} we describe the syntax and semantics of a pseudocode for a subset of the strict functional language Refal which will be used throughout this paper. We give also the operational semantics of the subset, defining its ``self-interpreter''. In Section \ref{sec:Specifying-Cache} we outline our approach for specifying non-deterministic systems by an example used through this paper.
In Section \ref{sec:Supercompilation} we shortly introduce an unfold-fold program transformation method known as Turchin's supercompilation that is used in our experiments.
We describe the strategy controlling the unfold-fold loop. The corresponding relation is a composition of Turchin's relation and a variant of the Higman-Kruskal relation. This composition plays a central role in verifying the safety properties of the cache coherence protocols' models via intermediate interpretation. 
In Section \ref{sec:Verif-Method-in-Action} we prove in a uniform way a number of properties of 
a huge number of 
the complicated configurations generated by specialization of the self-interpreter with respect to the given program modeling a cache coherence protocol. The argumentations given in the section are applicable for the whole series of the protocols mentioned in Section \ref{sec:Conclusion}. Developing the method of such argumentations is the main aim of the paper and \emph{both Proposition \ref{prop:Match} and the proof of this proposition are the main results of the paper}. The statement given in Proposition \ref{prop:Match} can be applied to a wide class of interpreters of Turing complete programming languages. 
This statement is a theoretical basis for the explanation why the approach suggested in the paper does succeed in verifying the safety properties of the series of the cache coherence protocols via intermediate interpretation. 
Finally, in Section \ref{sec:Conclusion} we report on some other experimental results obtained by using the approach, discuss the results presented in the paper, and compare our experiments with other ones done by existing methods.

\vspace{-5pt}
\section{An Interpreter for a Fragment of the SCP4 Object Language}\label{sec:Self-Interpreter}

A first interpreter we consider is an interpreter of a subset {${\cal L}$} of the SCP4 object language, which we aim to put in between the supercompiler SCP4 and programs 
modeling the cache coherence protocols to be verified. 
We will refer to this interpreter as a ``self-interpreter''.

\subsection{Language}\label{subsec:Self-Int-Language}

\begin{center}
    \begin{tabular}{rll}
$\expr{\var{prog}}$ ::= &  $\expr{\var{def_{1}}}$ $~\ldots~$ $\expr{\var{def_{n}}}$ & Program 
\\
$\expr{\var{def}}$ ::= &   
${\expr{\var{f}}\expr{\brackets{~\expr{\patt{\var{ps_{1}}}}}}}\Rightarrow$ $\expr{\var{exp_{1}}};$ $~\ldots;~$ 
${\expr{\var{f}}\expr{\brackets{~\expr{\patt{\var{ps_{n}}}}}}}\Rightarrow$ $\expr{\var{exp_{n}}};$
& Function Definition 
\\
$\expr{\var{exp}} ::=$ &	$\expr{\var{v}}$		& Variable\\
$|$	&	$\expr{\expr{\var{term}}}{\;:\;}\expr{\var{exp}}$ & $\expr{\var{Cons}}$ Application\\
$|$	&	$\expr{\var{f}}\textbf{(}$ $\expr{\var{exp_{1}}},$ $\expr{\var{...}},$ $\expr{\var{exp_{n}}}$ $\textbf{)}$ & Function Application\\
$|$	&	$\expr{\var{exp_{1}}}$ $\textbf{++}$ $\expr{\var{exp_{2}}}$ & $\expr{\var{Append}}$ Application\\
$|$	&	$\Nil$ & $\expr{\var{Nil}}$\\
$\expr{\var{term}}$ ::= & $s\textbf{.}name$ & Symbol-Type Variable \\ 
$|$ & $(exp)$   & Constructor Application \\ 
$|$ & $\sigma$  & Symbol\\ 
\\
$\expr{\var{ps}}$ ::= & $\expr{\var{p_{1}}},$ $\expr{\var{...}},$ $\expr{\var{p_{n}}}$ & Patterns\\ 
$\expr{\var{p}}$ ::= & $\expr{\var{v}}$ & \\ 
$|$	& ${s}\textbf{.}name{\;:\;}\expr{\expr{\var{p}}}$ & \\ 
$|$	&	$\expr{\brackets{\expr{\var{p_{1}}}}}{\;:\;}\expr{\var{p_{2}}}$ & \\
$|$	&	$\sigma{\;:\;}\expr{\var{p}}$ & \\
$|$	&	$\texttt{[]}$ & \\
\\		
$\expr{\var{v}}$ ::= & $s\textbf{.}name$ $|$ $e\textbf{.}name$ & Variable
  \end{tabular}
\end{center}

Programs in {${\cal L}$} are \emph{strict} term rewriting systems based on pattern matching. 

The rules in the programs are ordered from the top to the bottom to be matched. 
To be closer to Refal we use two kinds of variables: \emph{s.}variables range over \emph{symbols} (i.e., characters and identifiers, for example, $\texttt{'a'}$ and $\expr{\var{True}}$), while \emph{e.}variables range over the whole set of the \texttt{S}-expressions.\footnote{This fragment of Refal is introduced for the sake of simplicity.  The reader may think that the syntactic category $\expr{\var{exp}}$ of list expressions and the parentheses constructor are Lisp equivalents. Actually Refal does not include $\expr{\var{Cons}}$ constructor but, instead of $\expr{\var{Cons}}$, $\expr{\var{Append}}$ is used as an associative constructor. Thus the Refal data set is wider as compared with the Lisp data set: the first one is the set of finite sequences of \emph{arbitrary} trees, while the second one is the set of binary trees. See \cite{Turchin:Refal5} for details.\label{foot:append}}
Given a rule
${\expr{\var{l}}}\Rightarrow$ $\expr{\var{r}}$,
  any variable of  $\expr{\var{r}}$ should appear in ${\expr{\var{l}}}$.
 Each function ${\expr{\var{f}}}$ has a fixed arity, i.e., the arities of all left-hand sides of the rules of ${\expr{\var{f}}}$ and any expression  $\expr{\var{f}}\textbf{(}~\expr{\var{exp_{1}}},\; \expr{\var{...}},\; \expr{\var{exp_{n}}}~\textbf{)}$ 
 must equal the arity of ${\expr{\var{f}}}$. 
The parentheses constructor {$(\bullet)$} is used without a name. {$Cons$} constructor is used in infix notation and may be omitted.
The patterns in a function definition are not exhaustive. If no left-hand side of the function rules matches the values assigned to a call of the function, then executing the call is interrupted and its value is undefined. 
In the sequel, the expression set is denoted by {$\mathbb{E}$}; 
{$\mathbb{D}$} and {$\mathbb{S}$} stand for 
the data set, i.e., the patterns containing no variable, and the symbols set, respectively.
The name set of the functions of an arity {$n$} is denoted by {$\Fn{n}$} while {$\mathbb{F}$} stands for 
{$\bigcup_{n=0}^{\infty} \Fn{n}$}.
${\cal V}_e$ and ${\cal V}_s$  
stand for the \emph{e}- and \emph{s}-variable sets, respectively, and ${\cal V}$ denotes 
${\cal V}_e \cup {\cal V}_s$. 
For an expression {$\expr{\var{exp}}$} 
{${\cal V}_e(\expr{\var{exp}})$}, {${\cal V}_s(\expr{\var{exp}})$}, {${\cal V}(\expr{\var{exp}})$} denote the corresponding variable sets of {$\expr{\var{exp}}$}.
$\expr{\var{\mu_v(exp)}}$ denotes the multiplicity of $v \in {\cal V}$ in {$\expr{\var{exp}}$}, i.\,e., the number of all the occurrences of $v$ in {$\expr{\var{exp}}$}.
{$\expr{\var{exp}}$} is called passive if no function application occurs in {$\expr{\var{exp}}$} otherwise it is called an active expression.
${\cal T}$ stands for the term set, $\sigma$ stands for a symbol.
Given an expression $\expr{\var{exp}}$ and a variable substitution {$\theta$}, $\expr{\var{exp}}\theta$ stands for $\theta(\expr{\var{exp}})$.

\subsection{Encoding}\label{subsec:Encoding}

In our experiments considered in this paper the protocol program models have to be input values of the interpreter argument with respect to which the interpreter is specialized. Thus the program models should be encoded in the data set of the implementation language of the interpreter. 
The program models used in this paper are written in a fragment of the language described in Section \ref{subsec:Self-Int-Language}, where $\append$ constructor is not allowed and only unary functions may appear.

Now we have to define the corresponding encoding function denoted by the underline,
where the function $\mathfrak{A}$ groups the program rules belonging to the same function as it is 
shown in the second definition line.


\begin{center}
    \begin{tabular}{ll}
{$\expr{\underline{\expr{\var{prog}}}}$} $=$ {$\expr{\brackets{~~\underline{\expr{\enapp{A}{\expr{\var{prog}}}}}~~}};$} 
&  Program  \\
{$\expr{\underline{{\expr{\var{f}}{\,}{\{}\expr{\var{rules}}{\}~}\expr{\var{defs}}}}}$} $=$ {$\expr{\brackets{{\var{f}~~\underline{\expr{\var{rules}}}~}}}$}{\;:\;}{$\underline{\expr{\var{defs}}};$}
&  Function Definitions \\
{$\expr{\underline{{\expr{\var{rule;}}{~\;}\expr{\var{rules}}}}}$} $=$ {$\expr{\brackets{{~\underline{\expr{\var{rule}}}~~}}}$}{\;:\;}{$\underline{\expr{\var{rules}}};$}
& Rules \\
 {$\expr{\underline{\fapp{\expr{\var{f}}}{\var{pattern}}{\;\Rightarrow\;}\expr{\var{exp}}}}$} $=$ 
{$\expr{\brackets{{~~\underline{\expr{\var{pattern}}}~~}}}$} {\;:\;}\texttt{'='}{\;:\;} {$\expr{\brackets{{~~\underline{\expr{\var{exp}}}~~}}};$}
&  Rule \\
 {$\expr{\underline{\expr{\var{term}}{\;:\;}\expr{\var{exp}}}}$} $=$ 
{$\expr{{\underline{\expr{\var{term}}}}}$} {\;:\;} {$\expr{{\underline{\expr{\var{exp}}}}};$}
& 
 Here 
$\var{term}$ ${::=}$ $(\var{exp})$ $|$ $\var{s.name}$ $|$ $\sigma$ \\
%
 {$\expr{\underline{\expr{\brackets{\expr{\var{exp}}}}}}$} $=$ 
{$\expr{\brackets{{\texttt{'*'}~\;\underline{\expr{\var{exp}}}~}}};$}
\hspace{60pt}
{$\expr{\underline{\fapp{\expr{\var{f}}}{\var{exp}}{}}}$} $=$ 
{$\expr{\brackets{{\var{Call}~\;\var{f}~\;\underline{\expr{\var{exp}}}~}}};$}
 & Applications \\
%
 {$\expr{\underline{\var{e.name}{}}}$} $=$ 
{$\expr{\brackets{\expr{\var{Var}}~\;\var{\texttt{'}e\texttt{'}}~\;\expr{\var{name}}}};$}
\hspace{25pt}
 {$\expr{\underline{\var{s.name}{}}}$} $=$ 
{$\expr{\brackets{\expr{\var{Var}}~\;\var{\texttt{'}s\texttt{'}}~\;\expr{\var{name}}}};$}
 & Variables \\
%
 {$\expr{\underline{[]}}$} $=$ {$[];$}
\hspace{120pt}
 {$\expr{\underline{\var{\sigma}{}}}$} $=$ {$\var{\sigma};$} &
   {$\var{Nil}$} and Symbol\\

    \end{tabular}\label{Encoding}
\end{center}

Note that any pattern is an expression.

 Supercompiler SCP4 in its processing dealing with programs as  input data uses this 
encoding function and utilizes its properties.   
The image of $\mathbb{D}$ under the encoding is a proper subset of $\mathbb{D}$, i.\,e., $\underline{\mathbb{D}} \neq \mathbb{D}$. 
For example, $\expr{\brackets{\expr{\var{Var}}~\var{\texttt{'}s\texttt{'}}~\expr{\var{name}}}} \notin \underline{\mathbb{D}}$.
%


{\small 
\begin{figure}[htb]
\begin{center}
{\hrule height 0.8pt \medskip}
      %
%
%
\fsent{Int}{\patts{\brackets{\patt{\patt{\var{Call}}{\var{s.f}}}{\var{e.d}}}}{\var{e.P}}}{\fapp{Eval}{\args{\fapp{EvalCall}{\args{\args{\var{s.f}}{\var{e.d}}}{\var{e.P}}}}{\var{e.P}}}}
{\fnew}
%
%
%
\lfsent{Eval}{\patts{\brackets{\var{e.env}}{:\brackets{\patt{\patt{\var{Call}}{\var{s.f}}}{\var{e.q}}}{:\var{e.exp}}}}{e.P}}{\fapp{Eval}{\args{\fapp{EvalCall}{\threeargs{\var{s.f}}{\fapp{Eval}{\args{\brackets{\var{e.env}}{:\var{e.q}}}{\var{e.P}}}}{\var{e.P}}}}{\var{e.P}}}{\,\texttt{++}\,\fapp{Eval}{\args{\brackets{\var{e.env}}{:\var{e.exp}}}{\var{e.P}}}}}
%
%
%
\fsent{Eval}{\patts{\brackets{\var{e.env}}{:\brackets{\patt{\var{Var}}{\var{e.var}}}{:\var{e.exp}}}}{e.P}}
{\fapp{Subst}{\args{\var{e.env}}{\brackets{\app{\var{Var}}{\var{e.var}}}{}}}{}{\,\texttt{++}\,\fapp{Eval}{\args{\brackets{\var{e.env}}{:\var{e.exp}}}{\var{e.P}}}{}}}
%
%
%
\fsent{Eval}{\patts{\brackets{\var{e.env}}{:\brackets{\patt{\var{\texttt{'*'}}}{\var{e.q}}}{:\var{e.exp}}}}{e.P}}{\brackets{\app{\var{\texttt{'*'}}}{\fapp{Eval}{\args{\brackets{\var{e.env}}{:\var{e.q}}}{\var{e.P}}}}}{\;{:}\;\fapp{Eval}{\args{\brackets{\var{e.env}}{:\var{e.exp}}}{\var{e.P}}}{}}}
%
%
%
\fsent{Eval}{\patts{\brackets{\var{e.env}}{:\var{s.x}}{:\var{e.exp}}}{e.P}}
{\expr{\var{s.x}}{\;{:}\;\fapp{Eval}{\args{\brackets{\var{e.env}}{:\var{e.exp}}}{\var{e.P}}}{}}}
%
%
%
\fsent{Eval}{\patts{\brackets{\var{e.env}}{:\Nil}}{e.P}}{\Nil}
%
%
%
%
%
\shiftfsent{EvalCall}{\threepatts{\var{s.f}}{\var{e.d}}{\brackets{\patt{\var{Prog}}{\var{s.n}}}{}}}{\fapp{Matching}{\args{\args{\var{\False}}{\var{\Nil}}}\args{\fapp{LookFor}{\args{\var{s.f}}{\fapp{Prog}{\var{s.n}}}}}{\var{e.d}}}}
{\fnew}
%
%
\lfsent{Matching}{
 \patts{\patts{\var{\False}}{\var{e.old}}}
{\patts{\brackets{\brackets{\var{e.p}}{{\;:}\texttt{'='}{\;:\;}\brackets{\var{e.exp}}}}{{\;:\;}\var{e.def}}}{\var{e.d}}}
 }
{\fapp{Matching}{\args{\args{\fapp{Match}{\threeargs{\var{e.p}}{\var{e.d}}{\brackets{\texttt{[]}}}}}{\var{e.exp}}}{\args{\var{e.def}}{\var{e.d}}}}}
%
%
%
%
\fsent{Matching}{\patts{
 \patts{\brackets{\var{e.env}}}{\var{e.exp}}}
 {\patts{\var{e.def}}{\var{e.d}}}
   }
{\brackets{\var{e.env}}{:\var{e.exp}}}
%
%
%
\shiftfsent{Match}{
   \threepatts{\brackets{\var{Var}~\texttt{'e'}~\var{s.n}}}{\var{e.d}}{\brackets{\var{e.env}}{}}
   }
{{\fapp{PutVar}{\args{\brackets{\patt{\var{Var}}{\patt{\texttt{'e'}}{\var{s.n}}}}{\,:\;}\var{e.d}}{\brackets{\var{e.env}}{}}}{}}}
%
%
\shiftlfsent{Match}{
   \threepatts{\brackets{\var{Var}~\texttt{'s'}~\var{s.n}}{{\;:\;}\var{e.p}}}{\var{s.x}{\,:\,}\var{e.d}}{\brackets{\var{e.env}}{}}
   }
{\fapp{Match}{\threeargs{\expr{\var{e.p}}}{\expr{\var{e.d}}}{\fapp{PutVar}{\args{\brackets{\patt{\var{Var}}{\patt{\texttt{'s'}}{\var{s.n}}}}{\,:\;}\var{s.x}}{\brackets{\var{e.env}}{}}}{}} }}
%
%
%
\shiftlfsent{Match}{
   \threepatts{\brackets{\texttt{'*'}~\var{e.q}}{{\;:\;}\var{e.p}}}{\brackets{\texttt{'*'}~\var{e.x}}{{\;:\;}\var{e.d}}}{\brackets{\var{e.env}}{}}
   }
{\fapp{Match}{\threeargs{{\expr\var{e.p}}}{\expr{\var{e.d}}}{\fapp{Match}{\threeargs{\expr{\var{e.q}}}{\expr{\var{e.x}}}{\brackets{\var{e.env}}{}}}{}}}}
%
%
\shiftfsent{Match}{
\threepatts{\expr{\var{s.x}{{\;:\;}\var{e.p}}}}{\expr{\var{s.x}{{\;:\;}\var{e.d}}}}{\brackets{\expr{\var{e.env}}}{}}
   }
{\fapp{Match}{\threeargs{\expr{\var{e.p}}}{\expr{\var{e.d}}}{\brackets{\expr{\var{e.env}}}{}}}{}}  
%
\shiftfsent{Match}{
   \threepatts{\texttt{[]}}{\texttt{[]}}{\brackets{\var{e.env}}{}}
   }
{\brackets{\var{e.env}}}
%
\shiftfsent{Match}{
\threepatts{\var{e.p}}{\var{e.d}}{\var{e.fail}}}
{\var{\False}}
{\fnew}
\fsent{PutVar}{
   \patts{\var{e.assign}}{\brackets{\var{e.env}}{}}
   }
{\fapp{CheckRepVar}{\fapp{PutV}{\threeargs{\brackets{\var{e.assign}}}{\var{e.env}}{\texttt{[]}}}{}}}
%
%
%
\shiftlfsent{PutV}{
\threepatts{\brackets{\brackets{\var{Var}~\var{s.t}~\var{s.n}}{{\;:\;}\expr{\var{e.val}}}}}
{\brackets{\brackets{\var{Var}~\var{s.t}~\var{s.n}}{{\;:\;}\var{e.pval}}}{{\;:\;}\expr{\var{e.env}}}{}}{\var{e.penv}{}}
   }
{{{\expr{\brackets{\fapp{Eq}{\args{\var{e.val}}{\var{e.pval}}}}}}{\;:\brackets{\brackets{\var{Var}~\var{s.t}~\var{s.n}}{{\;:\;}\var{e.pval}}}{{\;:\;}{\expr{\var{e.env}}}}} }}
%
%
%
%
%
%
%
\shiftlfsent{PutV}{
 \threepatts{\brackets{\var{e.assign}}}{\brackets{\var{e.passign}}{{\;:\;}\var{e.env}}{}}
 {\var{e.penv}{}}
   }
{\fapp{PutV}{\threeargs{\brackets{\var{e.assign}}}{\var{e.env}}{\expr{\brackets{\expr{\var{e.passign}}}{\;:\;}\expr{\var{e.penv}}}}{}}}
%
%
%
%
\shiftfsent{PutV}{
   \threepatts{\brackets{\var{e.assign}}}{\texttt{[]}}{\var{e.penv}{}}
   }
{{{\expr{\brackets{\var{\True}}}}{\;:\expr{\brackets{\var{e.assign}}{\;:\;}\expr{\var{e.penv}}}}}}
%
%
%
%
{\fnew}
%
%
%
\shiftfsent{CheckRepVar}{
   \expr{\brackets{\var{\True}}{:}\var{e.env}}
   }
{\brackets{\var{e.env}}{}}
%
%
%
\shiftfsent{CheckRepVar}{
   \expr{\brackets{\var{\False}}{:}\var{e.env}}
   }
{\var{\False}}
{\fnew}
%
\fsent{Eq}{
   \patts{\var{s.x}{\;:\;}\var{e.xs}}{\var{s.x}{\;:\;}\var{e.ys}}
   }
{{\fapp{Eq}{\args{\var{e.xs}}{\var{e.ys}}}{}}}
%
%
%
\fsent{Eq}{
\patts{\brackets{\texttt{'*'}~\var{e.x}}{{\;:\;}\var{e.xs}}}{\brackets{\texttt{'*'}~\var{e.y}}{{\;:\;}\var{e.ys}}}
   }
{{\fapp{ContEq}{\threeargs{\fapp{Eq}{\args{\expr{\var{e.x}}}{\expr{\var{e.y}}}}{}}{\expr{\var{e.xs}}}{\expr{\var{e.ys}}}}{}}}
%
%
\fsent{Eq}{
   \patts{\texttt{[]}}{\texttt{[]}}
   }
{\var{\True}}
%
%
\fsent{Eq}{
\patts{\var{e.xs}}{\var{e.ys}}
   }
{\var{\False}}
%
%
\shiftfsent{ContEq}{
\threepatts{\var{\False}}{\var{e.xs}}{\var{e.ys}}}
{\var{\False}}
%
\shiftfsent{ContEq}{
\threepatts{\var{\True}}{\var{e.xs}}{\var{e.ys}}}
{{\fapp{Eq}{\args{\var{e.xs}}{\var{e.ys}}}{}}}
{\fnew}
%
%
\fsent{LookFor}{
   \patts{\var{s.f}}{\brackets{\var{s.f}{\;:\;}\var{e.def}}{{\;:\;}\var{e.P}}}
   }
{\var{e.def}}
%
%
\fsent{LookFor}{
   \patts{\var{s.f}}{\brackets{\var{s.g}{\;:\;}\var{e.def}}{{\;:\;}\var{e.P}}}
   }
{{\fapp{LookFor}{\args{\var{s.f}}{\var{e.P}}}{}}}
{\fnew}
%
%
\fsent{Subst}{
\patts{\brackets{\brackets{\var{Var}~\var{s.t}~\var{s.n}}{{\;:\;}\expr{\var{e.val}}}}{{\;:\;}\expr{\var{e.env}}}}{\brackets{\var{Var}~\var{s.t}~\var{s.n}}{}}
   }
{\var{e.val}}
%
%
\fsent{Subst}{
\patts{\var{\brackets{e.assign}}{{\;:\;}\expr{\var{e.env}}}}{\expr{\var{e.var}}}
   }
{{\fapp{Subst}{\args{\var{e.env}}{\var{e.var}}}{}}}

\medskip
{\hrule height 0.8pt}
\end{center} 
\caption{Self-Interpreter}\label{fig:Self-Interpreter}
\label{program}
\end{figure}
}

\subsection{The Interpreter}\label{subsec:Interpreter}

The self-interpreter used in the experiments is given in Figure \ref{fig:Self-Interpreter}. 

The entry point is of the form 
{${\fapp{Int}{\args{\expr{\brackets{\expr{\var{Call}}~\expr{\var{s.f}}~\expr{\var{e.d}}}}}{\expr{\var{e.P}}}}{}}$}. 
Here the first argument is the application constructor of the main function to be executed. The second argument provides the name of a program to be interpreted. The encoded source of the program will be returned by a function call of {$\expr{\var{Prog}}$} whenever it is asked by {$\expr{\var{EvalCall}}$}.  E.\,g.,
interpretation of program model Synapse N+1 given in Section \ref{subsec:Program-Model} starts with the following application 
{${\fapp{Int}{\args{\expr{\brackets{\expr{\var{Call}}~\expr{\var{Main}}~\underline{\expr{\var{d}_0}}}}}{\expr{\brackets{\expr{\var{Prog}}~\expr{\var{Synapse}}}}}}{}}$},
where {$\expr{\var{d}_0}$} is an input data given to program Synapse.
Due to the large size of the encoded programs we omit the definition of function {$Prog$}.

\clearpage

{${\fapp{EvalCall}{\args{\expr{\brackets{\expr{\var{Call}}~\expr{\var{s.f}}~\expr{\var{e.d}}}}}{\expr{\var{e.P}}}}{}}$}  
asks for the definition of function {$\expr{\var{s.f}}$}, calling {$ \expr{\var{LookFor}} $}, and initiates matching  the data given by {$\expr{\var{e.d}}$} against the patterns of the definition rules. In order to start this pattern matching, it imitates a fail happening in matching the data against a previous nonexistent pattern. 

Function {$\expr{\var{Matching}}$} runs over the definition rules, testing the result of matching the input data ({$\expr{\var{e.d}}$}) against the current pattern considered. In the case if the result is {$\expr{\var{\False}}$} the function calls function {$\expr{\var{Match}}$}, asking for matching the input data against the next rule pattern. The environment {$\expr{\var{e.env}}$} is initialized by {$\expr{\var{\texttt{[]}}}$}. 
If the pattern matching succeeds then 
{$\expr{\var{Matching}}$} returns  {$\expr{\brackets{\var{e.env}}{:}\var{e.exp}}$}, where expression {$\expr{\var{e.exp}}$} is the right-hand side of the current rule and the environment includes the variable assignments computed by the pattern matching. 
Function {$\expr{\var{Match}}$} is 
trying to match 
the input data given in its second argument, step by step, against the pattern given in its first argument. It computes the environment containing the variable substitution defined by the matching. 
If a variable 
 is encountered then function {$\expr{\var{PutVar}}$} calls {$\expr{\var{PutV}}$} looking for an assignment to the same variable and, if such an assignment exists, the function tests a possible coincidence of the new and old values assigned to the variable. 
The third rule of function {$\expr{\var{Match}}$} deals with the tree structure, calling this function twice.
 
 Function {$\expr{\var{Eval}}$} passes through an expression given in its second argument. The second {$\expr{\var{Eval}}$} rule deals with a variable and calls function {$\expr{\var{Subst}}$} looking the environment for the variable value and replacing the variable with its value.
 
 We intend to specialize interpreter {$\expr{\var{Int}}$} with respect to its second argument.
The corresponding 
 source code of the self-interpreter may be found in 
{\small \url{http://refal.botik.ru/protocols/Self-Int-Refal.zip}}.

\vspace{-10pt}
\section{Specifying Cache Coherence Protocols}\label{sec:Specifying-Cache}

\vspace{-5pt}
We  illustrate our method \cite{Lis_Nem:Programming, Lis_Nem:IJFCS08} for specifying 
non-deterministic systems by  an example used through this paper. 
The Synapse N+1 protocol definition given below is borrowed from \cite{Delzanno:Automatic}. 
The parameterized version of the protocol is considered and \emph{counting abstraction} is used in the specification. 
 The protocol has to react to five external non-deterministic events by updating its states being three integer counters. The initial value of counter {$\expr{\var{invalid}}$} is parameterized
(so it could be any positive integer), 
 while the other two counters are initialized by zero. 
 The primed state names stand for the updated state values. The empty updates mean that nothing happened.

\begin{tabular}{ll}
  (rh) & \psent{dirty + valid \ge 1}{} \hspace{60pt}\hspace{55pt}(wh1)\hspace{10pt} \psent{dirty \ge 1}{} \\
  (rm) & \psent{invalid \ge 1}{dirty^\prime = 0, valid^\prime = valid + 1, invalid^\prime = invalid + dirty - 1} \\
 (wh2) & \psent{valid \ge 1}{valid^\prime = 0, dirty^\prime = 1, invalid^\prime = invalid + dirty + valid - 1} \\
 (wm) & \psent{invalid \ge 1}{valid^\prime = 0, dirty^\prime = 1, invalid^\prime = invalid + dirty + valid - 1}
\end{tabular}

\vspace{-10pt}
\paragraph{Specification of Safety Properties}\label{subsubsec:Spec-Safety-Prop}
Any state reached by the protocol should not satisfy any of the two following properties:
%
%
  (1)~~  \pprop{invalid \ge 0, dirty \ge 1, valid \ge 1};
  (2)~~  \pprop{invalid \ge 0, dirty \ge 2, valid \ge 0}.

\vspace{-5pt}
\subsection{Program Model of the Synapse N+1 Cache Coherence Protocol}\label{subsec:Program-Model}

The program model of Synapse N+1 protocol is given in Figure \ref{fig:Synapse}.
The idea behind the program specifications modeling the reactive systems is given in Introduction  \ref{sec:Introduction} above. The \emph{finite} stream of events is modeled by a $\expr{\var{time}}$ value. The time ticks
are labeled by the events.  The counters' values are specified in the unary notation. The unary addition is directly defined by function $\expr{\var{Append}}$, i.e., without referencing to the corresponding macros. 
Function $\expr{\var{Loop}}$ exhausts 
the event stream, step by step, and calls for $\expr{\var{Test}}$ verifying the safety property required from the protocol. Thus function $\expr{\var{Main}}$ is a predicate.
Note that given input values the partial predicate terminates since the event stream is finite. The termination is normal, if the final protocol state asked by the input stream is reachable one, otherwise it is abnormal.

\begin{figure}[htb]
\begin{center}
{\hrule height 0.8pt \medskip}

\fsent{Main}{\expr{\brackets{\var{e.time}}}{\,:\,}\brackets{\expr{\var{e.is}}}}
{\fapp{Loop}{\brackets{\var{e.time}}{\,:\,}\expr{\brackets{\var{Invalid}~\var{I}~\var{e.is}}}{\,:\,}\expr{\brackets{\var{Dirty}~}}{\,:\,}\expr{\brackets{\var{Valid}~}}}}
%
{\fnew}
\lfsent{Loop}{\brackets{\texttt{[]}}{\,:\,}{\expr{\brackets{\var{Invalid}~\var{e.is}}}}{\,:\,}{\expr{\brackets{\var{Dirty}~\var{e.ds}}}}{\,:\,}{\expr{\brackets{\var{Valid}~\var{e.vs}}}}}
{\fapp{Test}{\brackets{\var{Invalid}~\var{e.is}}{\,:\,}\brackets{\var{Dirty}~\var{e.ds}}{\,:\,}\brackets{\var{Valid}~\var{e.vs}}}}
\lfsent{Loop}{\brackets{\var{s.t}{\,:\,}\var{e.time}}{\,:\,}\expr{\brackets{\var{Invalid}~\var{e.is}}}{\,:\,}\expr{\brackets{\var{Dirty}~\var{e.ds}}}{\,:\,}\expr{\brackets{\var{Valid}~\var{e.vs}}}}
{\fapp{Loop}{\brackets{\var{e.time}}{\,:\,}\fapp{Event}{\var{s.t}{\,:\,}\brackets{\var{Invalid}~\var{e.is}}{\,:\,}\brackets{\var{Dirty}~\var{e.ds}}{\,:\,}\brackets{\var{Valid}~\var{e.vs}}}}}
%
%
{\fnew}
\lfsent{Event}{\var{rm}{\,:\,}\expr{\brackets{\var{Invalid}~\var{I}~\var{e.is}}}{\,:\,}\expr{\brackets{\var{Dirty}~\var{e.ds}}}{\,:\,}\expr{\brackets{\var{Valid}~\var{e.vs}}}}
{\brackets{\var{Invalid}~\fapp{Append}{\brackets{\var{e.ds}}{\,:\,}\brackets{\var{e.is}}}}{\,:\,}\brackets{\var{Dirty}~}{\,:\,}\brackets{\var{Valid}~\var{I}~\var{e.vs}}}
\lfsent{Event}{\var{wh2}{\,:\,}\expr{\brackets{\var{Invalid}~\var{e.is}}}{\,:\,}\expr{\brackets{\var{Dirty}~\var{e.ds}}}{\,:\,}\expr{\brackets{\var{Valid}~\var{I}~\var{e.vs}}}}
{\brackets{\var{Invalid}~\fapp{Append}{\brackets{\var{e.vs}}{\,:\,}\brackets{\fapp{Append}{\brackets{\var{e.ds}}{\,:\,}\brackets{\var{e.is}}}}}}{\,:\,}\brackets{\var{Dirty~\var{I}}}{\,:\,}\brackets{\var{Valid~}}}
\lfsent{Event}{\var{wm}{\,:\,}\expr{\brackets{\var{Invalid}~\var{I}~\var{e.is}}}{\,:\,}\expr{\brackets{\var{Dirty}~\var{e.ds}}}{\,:\,}\expr{\brackets{\var{Valid}~\var{e.vs}}}}
{\brackets{\var{Invalid}~\fapp{Append}{\brackets{\var{e.vs}}{\,:\,}\brackets{\fapp{Append}{\brackets{\var{e.ds}}{\,:\,}\brackets{\var{e.is}}}}}}{\,:\,}\brackets{\var{Dirty~\var{I}}}{\,:\,}\brackets{\var{Valid}{~}}}
%
%
\shiftfsent{Append}{\expr{\brackets{\texttt{[]}}}{\,:\,}\expr{\brackets{\var{e.ys}}}}
{\var{e.ys}}
%
\shiftfsent{Append}{\expr{\brackets{\var{s.x}{\,:\,}\var{e.xs}}}{\,:\,}\expr{\brackets{\var{e.ys}}}}
{\var{s.x}{\,:\,}\fapp{Append}{\brackets{\var{e.xs}}{\,:\,}\brackets{\var{e.ys}}}}
%
%
{\fnew}
\fsent{Test}{\expr{\brackets{\var{Invalid}~\var{e.is}}}{\,:\,}\expr{\brackets{\var{Dirty}~\var{I}~\var{e.ds}}}{\,:\,}\expr{\brackets{\var{Valid}~\var{I}~\var{e.vs}}}}
{\var{False}}
\fsent{Test}{\expr{\brackets{\var{Invalid}~\var{e.is}}}{\,:\,}\expr{\brackets{\var{Dirty}~\var{I}~\var{I}~\var{e.ds}}}{\,:\,}\expr{\brackets{\var{Valid}~\var{e.vs}}}}
{\var{False}}
\fsent{Test}{\expr{\brackets{\var{Invalid}~\var{e.is}}}{\,:\,}\expr{\brackets{\var{Dirty}~\var{e.ds}}}{\,:\,}\expr{\brackets{\var{Valid}~\var{e.vs}}}}
{\var{True}}

\medskip
{\hrule height 0.8pt}
\vspace{-5pt}

\end{center} 

\caption{Model of the Synapse N+1 cache coherence protocol}\label{fig:Synapse}
\end{figure}


\vspace{-5pt}
\section{On Supercompilation}\label{sec:Supercompilation}

\vspace{-5pt}
In this paper we are interested in one particular approach in program transformation and specialization, known as supercompilation\footnote{From \emph{super}vised \emph{compilation}.}.
Supercompilation is a powerful semantics-based program transformation
technique~\cite{SGJ:96,Tur:86} having a long history well back to the
1960-70s, when it was proposed by V.~Turchin. The main idea behind a supercompiler is to observe the
behavior of a functional program $p$ running on a \emph{partially} defined input 
with the aim to define a program, which would be equivalent to the original one (on the domain of the latter), but having improved properties. 
Given a program and its parameterized entry point, supercompilation is performed by an \emph{unfold-fold cycle} unfolding this entry point to a potentially infinite tree of all its possible computations. 
It reduces the redundancy that could be present in the original program. 
It folds the tree into a finite graph of states and transitions  between possible parameterized
configurations of the computing system. 
And, finally, it analyses global properties of the graph and
specializes this graph with respect to these properties (without additional
unfolding steps).\footnote{See also Appendix to the extended version of this paper \cite{Lis_Nem:Veri-via-Inters-arXiv}.} The resulting program definition is constructed solely based on the meta-interpretation of the source program rather than by a (step-by-step)
transformation of the program.  
The result of supercompilation may be a specialized version of the original program, taking into account the properties of partially known arguments, or just a re-formulated, and sometimes more efficient, equivalent program (on the domain of the original).

Turchin's ideas have been studied by a number of authors for a long time and
have, to some extent, been brought to the algorithmic and implementation stage
\cite{NT:00}. From the very beginning the development of supercompilation has
been conducted mainly in the context of the programming language Refal
\cite{N:03, Nemytykh:SCP4book, Nem_Pin_Tur, Turchin:Refal5}.
A number of model supercompilers for subsets of functional languages based on Lisp data were implemented with the aim of formalizing some aspects of the supercompilation algorithms 
\cite{Jonsson_Nordlander, Klyuchnikov:HOSC, SGJ:96}. 
The most advanced supercompiler for Refal is SCP4 \cite{N:03, Nemytykh:SCP4book, NT:00}. 

The verification system VeriMAP \cite{VeriMAP:2014} by E.\,De~Angelis et al. \cite{FioPettProSen:Ver-IterSpec14, FioPettProSen:Ver-Linear-Fib15} uses nontrivial properties of integers recognized by both CLP built-in predicates and external SMT solvers. 
We  use also a nontrivial property of the configurations. The property is the associativity of the built-in append function \texttt{++} supported by 
the supercompiler SCP4 itself\footnote{As well as by the real programming language in terms of which the experiments described in this paper were done.}, rather than by an external solver.

\subsection{The Well-Quasi-Ordering on {${\mathbb{E}}$}}\label{subsec:Higman-Kruskal}

 The following relation is a variant of the Higman-Kruskal relation and is a well-quasi-ordering \cite{Higman, Kruskal} (see also \cite{Leuschel:98}).

\numberwithin{equation}{section}

\begin{definition}
The homeomorphic embedding relation $\proptoeq$ is the smallest transitive relation on $\mathbb{E}$ satisfying the following properties, 
where $f \in \Fn{n},\; \alpha, \beta, \tau, s, t, t_{1}, \ldots, t_{n} \in {\mathbb E}$
and {$\alpha, \beta, \tau \in {\cal T}$}.
\\
%
 {\small (1)} ${\forall x,y \in  {\cal V}_e.\; x \proptoeq y, \forall u,v \in  {\cal V}_s.\; u \proptoeq v;}
 $ 
%
 \hspace{35pt}{\small (2)} ${\Nil \proptoeq t,\; t \proptoeq t,\; t \proptoeq f(\;{t_{1}},\; {\ldots},\; t,\ldots, {t_{n}}\;),\; 
t \proptoeq (t),\; 
t \proptoeq \alpha{\,{:}\,}t;} $
  \\
%
%
 {\small (3-4)} $\text{if~\;} {s \proptoeq {\zb}t \text{~and~} \alpha \proptoeq {\zb}\beta, \text{~then~both~}  
{(}s{)} \proptoeq {(}t{)},   
\alpha{\texttt{:}\,}s \proptoeq \,\beta{\,{:}\;}t} 
%
 \text{~and~}
{f(\;t_{1},\; \ldots,\; s,\ldots, t_{n}\;) \proptoeq {\zb}f(\;t_{1},\; {\ldots},\; t,\ldots, t_{n}\;)}.  
 $ 

\end{definition}\label{def:emb}

Note that the definition takes into account function {$\expr{\var{append}}$}, since its infix notation {$\expr{\var{exp_1}}\; \append\; \expr{\var{exp_2}}$} stands for {$\texttt{ append}(~\expr{\var{exp_1}},\; \expr{\var{exp_2}}~)$}.
We use relation $\proptoeq$ modulo associativity of \texttt{\;++\;} and the following equalities 
{${\expr{\var{term}}{\;:\;}\expr{\var{exp_1}}} = {\expr{\var{term}}~\append~\expr{\var{exp_1}}}
$}, {${\expr{\var{exp}} ~\append~ \Nil} = \expr{\var{exp}}$} and {${\Nil ~\append~ \expr{\var{exp}}} = \expr{\var{exp}}$}.

Given an infinite sequence of expressions $t_1, \ldots, t_n, \ldots$, relation $\proptoeq$ is relevant to approximation of loops increasing the syntactical structures 
in the sequence; or in other words to looking for the regular similar cases of mathematical induction on the structure of the expressions. That is to say the cases, which allow us to refer one to another by a step of the induction. 
An additional restriction separates the basic cases of the induction from the regular ones. The restriction is:  
{$
 \forall \sigma\; . \texttt{(\Nil)} \notpropto \texttt{(}\sigma\texttt{)} \;\&\;
 \forall v \in {\cal V}_s. \texttt{(\Nil)} \notpropto \texttt{(}v\texttt{)}
$}.

We impose this restriction on the relation $\proptoeq$ modulo the equalities above and denote 
the obtained relation as $\precceq$. It is
easy to see that such a restriction does not violate the quasi-ordering 
property. Note that the restriction may be varied in the obvious way, but for our experiments its simplest case given above is used to control generalization and has turned out to be sufficient. 
In the sequel,  $t_1 \prec t_2$ stands for the following relation $t_1 \precceq t_2$ and $t_1 \neq t_2$, which is also transitive.

\begin{definition}\label{def:configuration}
A parameterized configuration is a finite sequence of the form\\
\emph{$\texttt{let}~ \expr{\var{e.h} ~\texttt{=}~ 
                  \fapp{f_1}{\threeargs{\expr{\var{exp_{11}}}}{~\ldots~}{\expr{\var{exp_{1m}}}}}{}
                   ~~\texttt{in}}~~
                   \ldots\;~~ 
        \texttt{let}~ \expr{\var{e.h} ~\texttt{=}~ 
                  \fapp{f_k}{\threeargs{\expr{\var{exp_{k1}}}}{~\ldots~}{\expr{\var{exp_{kj}}}}}{}                          ~~\texttt{in}~~
        \expr{\var{exp_{n+1}}}}
$},
where $\expr{\var{exp_{n+1}}}$ is passive, for all $i>1$ $\expr{\var{\mu_{\expr{\var{e.h}}}(\fapp{f_i}{\ldots})}} = \expr{\var{\mu_{\expr{\var{e.h}}}(exp_{n+1})}} = 1$, and $\expr{\var{\mu_{\expr{\var{e.h}}}(\fapp{f_1}{\ldots})}} = 0$; 
for all $i$ and all $j$ variable {$\expr{\var{e.h}}$} does not occur in any function application being a sub-expression of $\expr{\var{exp_{ij}}}$. 
In the sequel, we refer to such a function application $\fapp{f_i}{\ldots}$ given explicitly in the configuration as an upper function application.
\end{definition}\label{def:config}

The configurations  represent 
the function application stacks, in which all constructors' applications not occurring in arguments of 
 the upper function applications are moved to the rightmost expressions. 
 Every expression of ${\cal L}$ can be rewritten into an equivalent composition of the configurations connected with $\var{let}$-construct (see Section \ref{sub:Meta-Reasoning} for an example).
 Here the \emph{append} {$\append$} is treated as a complex constructor\footnote{I.e., we use nontrivial properties of configurations containing the $\expr{\var{append}}$ \texttt{++}. See the remark 
 given in the footnote on p.\,\pageref{foot:append}.}
, rather than a function. The rightmost expression is the bottom of the stack. Since the value of {$\expr{\var{e.h}}$} is reassigned in each \texttt{let} in the stack, for brevity sake, we use the following presentation of the configurations: \\
{$\fapp{f_1}{\threeargs{\expr{\var{exp_{11}}}}{~\ldots~}{\expr{\var{exp_{1m}}}}},\;~
                   \ldots\;,~ 
  \fapp{f_k}{\threeargs{\expr{\var{exp_{k1}}}}{~\ldots~}{\expr{\var{exp_{kj}}}}},\;~                        \expr{\var{exp_{n+1}}}
$},
where variable $\expr{\var{e.h}}$ is replaced with bullet {$\bullet$}. I.e., the bullet is just a placeholder. The last expression may be omitted if it equals {$\bullet$}.
An example follows:  
{$
\expr{\fapp{f}{\var{a}{\;:\,}\var{e.xs}\append \var{e.ys}},}~~
\expr{\fapp{g}{\threeargs{\bullet \append \var{e.ys}}{(\var{Var}~ b~c)}{\Nil}},}~~ 
\expr{\fapp{f}{\var{s.x}{\;:\,}\bullet},}~ 
\expr{\var{s.x}{\;:\,} \bullet\append \fapp{t}{\var{s.x}{\;:\,}\var{e.zs}},}~~\bullet. 
$}

\subsection{The Well-Disordering on Timed Configurations}\label{subsec:Turchin}

Let a program to be specialized and a path starting at the root of the tree unfolded by the unfold-fold loop widely used in program specialization be given. The vertices in the path are labeled by the program parameterized configurations. These configurations form a sequence. Given a configuration from such a sequence and a function application from the configuration, we label the application by the time when it is generated by the unfold-fold loop. Such a labeled function application is said to be a timed application. A configuration is said to be timed if all upper function applications in the configuration are timed. 
Given a timed configuration, all its timed applications have differing time-labels. 
Given two different configurations $C_1, C_2$, if the unfold-fold loop copies an upper function application from $C_1$ and uses this copy in $C_2$, then $C_1, C_2$ share this timed application. 
In the sequel, a sequence of the timed configurations generated by the unfold-fold loop is also called just a path.   
In this section we define a binary relation $\tur$ on the timed configurations in the path. The relation is originated from V.\,F.~Turchin \cite{Turchin:Generalization88} (see also \cite{Nemytykh:SCP4book, AntoninaNepeivoda:Sbornik2014, AntoninaNepeivoda:VPT2016}). It is \emph{not} transitive\footnote{See Appendix to the extended version of this paper \cite{ Lis_Nem:Veri-via-Inters-arXiv} for an example 
demonstrating the nontransitivity of relation $\tur$.}, but like the well-quasi-ordering it satisfies the following crucial property used by supercompilation to stop the loop unfolding the tree.
For any infinite path $C_1, C_2,\; \ldots,\; C_n,\; \ldots$ there exist two timed configurations $C_i, C_j$ such that  $i < j$ and $C_i \tur C_j$ (see \cite{Turchin:Generalization88, AntoninaNepeivoda:VPT2016}). For this reason we call relation $\tur$ a well-disordering relation. In the sequel, the time-labels are denoted with subscripts.

\begin{definition}\label{def:Turchin}

Given a  sequence of timed configurations  $C_1,\; \ldots\; C_n,\; \ldots$;  $C_i$  and $C_j$ are elements of  the sequence such that $i < j$ and 
$ C_i \; ::=\; f_{t_1}^1(\;\ldots\;),\, \ldots,\, f_{t_k}^k(\;\ldots\;), \expr{\var{exp_1}}$, 
$ C_j\; ::=\; g_{\tau_1}^1(\;\ldots\;),\, \ldots,\, g_{\tau_m}^m(\;\ldots\;), \expr{\var{exp_2}}$, 
where  $f_{t_s}^s$ and $g_{\tau_q}^q$, $1 \le s \le k$ and $1 \le q \le m$, stand for function names $f^s$, $g^q$ labeled with times ${t_s}$ and  ${\tau_q}$, respectively, and $\expr{\var{exp_1}}, \expr{\var{exp_2}}$ are passive expressions.

If $k \le m$, $\delta = m-k$ and $\exists l\,.\, (1 < l \le k)$ such that $\forall s\,.\, (0 \le s \le k-l)$ 
$f_{t_{l+s}}^{l+s} = g_{\tau_{\delta+l+s}}^{\delta+l+s}$ (i.e., $f^{l+s} = g^{\delta+l+s}$ and $ {t_{l+s}} = {\tau_{\delta+l+s}}$ hold),  $f_{t_{l-1}}^{l-1} \neq g_{\tau_{\delta+l-1}}^{\delta+l-1}$, and 
$\forall s\,.\, (0 < s < l)$ $f_{t_{s}}^s \simeq g_{\tau_{s}}^s$ (i.e., $f^s = g^s$),
then $C_i \tur C_j$. 

We say that configurations $C_i$, $C_j$ are in Turchin's relation $C_i \tur C_j$.
This longest coincided suffix of the configurations are said to be the context, while the parts equal one to another modulo their time-labels are called prefixes of the corresponding configurations.  
\end{definition}

The idea behind this definition is as follows. The function applications in the context never took a part in computing the configuration $C_j$, in this segment of the path, while any upper function application in the prefix of $C_i$ took a part in computing the configuration $C_j$. Since the prefixes of $C_i, C_j$ coincide modulo their time-labels, these prefixes approximate a loop in the program being specialized. The prefix of $C_i$ is the entry point in this loop, while the prefix of $C_j$ initiates the loop iterations.
The common context approximates computations after this loop.
Note that Turchin's relation does not impose any restriction on the arguments of the function applications in $C_i, C_j$.

For example, consider the following two configurations 
$C_1 ~::= ~ f_4(\ldots),\, f_3(\ldots),\, g_2(\ldots),\, t_1(\ldots),\, \bullet$ ~and~
$C_2 ~::= ~ f_{10}(\ldots),\, f_7(\ldots),\, g_5(\ldots),\, ~\ldots,~ t_1(\ldots),\, \bullet$ , then $C_1 \tur C_2$ holds. Here the context is $t_1(\ldots)$, 
the prefix of $C_1$ is $f_4(\ldots),\, f_3(\ldots),\, g_2(\ldots)$, and 
the prefix of $C_2$ is $f_{10}(\ldots),\, f_7(\ldots),\, g_5(\ldots)$, 
where the subscripts of the application names stand for the time-labels.
See also Appendix to the extended version of this paper \cite{ Lis_Nem:Veri-via-Inters-arXiv} for a detailed example regarding Turchin's relation.

\vspace{-5pt}
\subsection{The Strategy Controlling the Unfolding Loop}\label{subsec:Strategy}

Now we describe the main relation controlling the unfold-fold loop. That is to say, given a path starting at the root of the unfolded tree, and two timed configurations $C_1, C_2$ in the path such that $C_1$ was generated before $C_2$, this relation stops the loop unfolding the tree and calls 
 the procedures responsible for folding this path. These tools, firstly, attempt to fold $C_2$ by a previous configuration and, if that is impossible, then attempt to generalize this configuration pair.  
The relation is a composition of relations $\tur$ and $\precceq$. It is denoted with $\turprec$ and is  a well-disordering (see \cite{Turchin:Generalization88, AntoninaNepeivoda:VPT2016}).

Thus we are given two timed configurations $C_1, C_2$ from a path, such that $C_1$ is generated before $C_2$, and $C_2$ is the last configuration in the path. If relation $C_1 \tur C_2$ does not hold, then the unfold-fold loop unfolds the current configuration $C_2$ and goes on. In the case relation $C_1 \tur C_2$ holds, these configurations are of the forms (see Section \ref{subsec:Turchin} for the 
notation used below): \\
$C_1 \; =\; f_{t_1}^1(\;\ldots\;),\, \ldots,\, f_{t_{l-1}}^{l-1}(\;\ldots\;), 
   f_{t_l}^l(\;\ldots\;), \ldots,\, f_{t_k}^k(\;\ldots\;), \expr{\var{exp_1}}$, \\
$C_2\; =\; f_{\tau_1}^1(\;\ldots\;),\, \ldots,\, f_{\tau_{l-1}}^{l-1}(\;\ldots\;),
   g_{\tau_l}^l(\;\ldots\;),\, \ldots,\, g_{\tau_{m}}^{m}(\;\ldots\;),\ f_{t_l}^l(\;\ldots\;), \ldots,\, f_{t_k}^k(\;\ldots\;), \expr{\var{exp_2}}$, 
where the context starts at $f_{t_l}^l(\;\ldots\;)$.
Let $C_i^{p}$ stand for the prefix of $C_i$, and $C_i^{c}$ stand for the context of $C_i$ followed by $\expr{\var{exp_i}}$.

Now we compare the prefixes as follows. 
If 
there exists $i$
$(1 \le i < l)$ such that $f_{t_i}^i(\;\ldots\;) \precceq f_{\tau_i}^i(\;\ldots\;)$ does not hold, then $C_2$ is unfolded and the unfold-fold loop goes on. 
Otherwise, 
the sub-tree rooted in $C_1$ is removed and the specialization task defined by the $C_1$ is  decomposed into the two specialization tasks corresponding to $C_1^{p}$ and $C_1^{c}$. 
Further the 
attempts to fold $C_2^{p}$ by $C_1^{p}$ and $C_2^{c}$  by $C_1^{c}$ 
do work. If some of these attempts fail, then the corresponding configurations are generalized. Note that the context may be generalized despite the fact that it does not take a part in computing the current configuration $C_2$, since a narrowing of the context parameters may have happened.

A program configuration is said to be a transitive configuration if one-step unfolding of the configuration results in a tree containing only the vertices with at most one outgoing edge. 
For example, any function application of the form $f(d_1, \ldots, d_n)$, where any $d_i \in \mathbb{D}$,  is transitive.
For the sake of simplicity, in the experiments described in this paper, the following strategy is used. The unfold-fold loop skips all transitive configurations encountered and removes them from the tree being unfolded.
In the sequel, we refer to the strategy described in this section, including relation $\turprec$, as the $\turprec$-strategy.

\vspace{-5pt}
\section{Indirect Verifying the Synapse N+1 Program Model}\label{sec:Verif-Method-in-Action}

\vspace{-5pt}
In this section we present an application of our program verification method based on supercompilation of intermediate interpretations. In general the method may perform a number of program specializations\footnote{I.\,e., the iterated ordinary supercompilation, which does not use any intermediate interpretation,  of the residual program produced by one indirect verification.}, but all the cache coherence protocol program models that we have tried to verify by supercompiler SCP4 require at most two specializations.

Given a program partial predicate modeling both a cache coherence protocol and a safety property  of 
the protocol, we use supercompilation aiming at moving the property hidden in the program semantics to a simple syntactic property of the residual program generated by supercompilation, i.e., this syntactic property should be easily recognized. In the experiments discussed in this paper we hope the corresponding residual programs will include no operator {$\expr{\var{return}~\var{False}};$}. 
Since the original direct program model terminates on any input data (see Section \ref{subsec:Program-Model}), this property means that the residual predicate never returns {$\expr{\var{False}}$} and always {$\expr{\var{True}}$}. 
Thus we conclude the original program model satisfies the given safety property. In the terms of functional language {${\cal L}$} presented in Section \ref{subsec:Self-Int-Language} the corresponding syntactic property is \emph{``No rule's right-hand side contains identifier {$\expr{\var{False}}$}''}.\footnote{Actually {$\expr{\var{False}}$} is never encountered at all in any residual program generated by repeated launching the supercompiler SCP4 verifying the cache coherence protocol models considered in this paper. I.e., the property is simpler than the formulated one.

Given a safety property required from a protocol, in order to look for witnesses violating the property, the method above can be extended by deriving $\var{False}$ by unfolding, using a specializer in an interactive mode. See \cite{Lis_Nem:Protocols07, Lis_Nem:TAP08, Lis_Nem:MissCannVPT2014, ANepeivoda:PingPong} for examples of bugged protocols and the corresponding witnesses constructed by means of the supercompiler SCP4.
}

We can now turn to the program modeling the Synapse N+1 protocol given in Section \ref{subsec:Program-Model}.
In order to show that the Synapse program model is safe, below we specialize the self-interpreter {$\expr{\var{Int}}$} (see Section \ref{subsec:Interpreter}) with respect to the Synapse program model  rather than the program model itself. 
Since program Synapse terminates, the self-interpreter terminates when it interprets any call of the {$\expr{\var{Main}}$} entry function of Synapse. 
Since Synapse is a partial predicate, the calls of the form 
{\small $\fapp{Int}{\hspace{-2pt}\args{\expr{\brackets{\expr{\var{Call}}~\expr{\var{Main}}~\expr{\var{e.d}}}}}{\expr{\brackets{\expr{\var{Prog}}~\expr{\var{Synapse}}}}}\hspace{-2pt}}{}$}, 
where {$e.d$} takes any data, define  the same partial predicate. Hence, the self-interpreter restricted to such calls is just another program model of protocol Synapse N+1. This indirect program model is much more complicated as compared with the direct model. 
We intend now to show that supercompiler SCP4 \cite{N:03, Nemytykh:SCP4book, NT:00} is able to verify this model.
Thus our experiments show potential capabilities of the method for verifying the safety properties of the functional programs modeling some complex non-deterministic parameterized systems. In particular, the experiments can also be considered as a partial verification of the intermediate interpreter used. In other words, verifying the interpreter with respect to a set of the interpreted programs that specify the cache coherence protocols.
This specialization by supercompilation is performed by following the usual \emph{unfold-fold cycle} controlled by the $\turprec$-strategy described in Section \ref{subsec:Strategy}.
Note that this program specification includes both the function call and constructor application stacks, where the size of the first one is uniformly bounded on 
the value of 
the input parameter while the second one is not.

We start off by unfolding the 
 initial configuration
{$
{\fapp{Int}{\args{\expr{\brackets{\expr{\var{Call}}~\expr{\var{Main}}~\expr{\var{e.d}}}}}{\expr{\brackets{\expr{\var{Prog}}~\expr{\var{Synapse}}}}}}{}}
$}, 
where the value of {$\expr{\var{e.d}}$} is unknown.
The safety property will be proved if supercompilation is able to recognize all rules of the interpreted program model, containing  the {$\expr{\var{False}}$} identifier, as being unreachable from this 
initial configuration.

In our early work \cite{Lis_Nem:IJFCS08} we have given a formal model of the verification procedure above by supercompilation. Let a program model and its safety property be given as described above, i.e., a partial program predicate. Given an initial parameterized configuration of the partial predicate, it has been shown that the unfold-fold cycle may be seen as a series of proof attempts by structural induction over the program configurations encountered during supercompilation aiming at verification of the safety property. Here the initial configuration specifies the statement that we have to prove. 
There are too many program configurations generated by the unfold-fold cycle starting with the initial configuration given above and the self-interpreter configurations are very large. As a consequence it is not possible to consider all the configurations in details. We study the configurations' properties being relevant to the proof attempts and the method for reasoning on such properties.

\subsection{On Meta-Reasoning}\label{sub:Meta-Reasoning}

Let a program $P_0$ written in ${\cal L}$ and a function application $f(d_0)$, where $d_0 \in \mathbb{D}$ is its input data, be given. Let $\expr{\var{\Prog}}$ stand for expression $(\expr{\var{Prog}~\; \var{N_{P_0}}})$, where $\expr{\var{N_{P_0}}}$ is the program name.
The unfolding loop standing alone produces a computation path $C_0, C_1, \ldots, C_n, \ldots$ starting off $f(d_0)$. If $f(d_0)$ terminates then the path is finite $C_0, C_1, \ldots, C_k$. In such a case, for any $0 \le i < k$ $C_i$ is a configuration not containing parameters, while $C_k$ is either a passive expression, if partial function $f$ is defined on the given input data, or the abnormal termination sign $\bot$ otherwise. 
The unfolding iterates function $\expr{\var{step}}(\cdot)$ such that  $\expr{\var{step}}(C_i) = C_{i+1}$.

Now let us consider the following non-parameterized configuration 
$K_0 = \expr{\var{Eval}}( (\Nil){\;:\;}\underline{f(\; d_0 \;)}, \expr{\var{\Prog}} )$ 
of the self-interpreter. If $f(d_0)$ terminates then the loop unfolding the configuration $K_0$ results in the encoded passive configuration produced by the loop unfolding $f(d_0)$.
\\
\text{\hspace{20pt}}
$
K_1 = \fapp{step}{K_0} = 
\fapp{Eval}{\args{\fapp{EvalCall}{\threeargs{\var{\underline{f}}}{\fapp{Eval}{\args{\brackets{\var{\Nil}}{:\var{\underline{d_0}}}}{\var{\Prog}}}}{\var{\Prog}}}}{\var{\Prog}}}{\,\append\,\fapp{Eval}{\args{\brackets{\var{\Nil}}{:\var{\Nil}}}{\var{\Prog}}}}
$

Expression $K_1$ is not a configuration. According to the strategy described in Section \ref{subsec:Interpreter} the unfolding has to decompose expression $K_1$ in a sequence of configurations connected by the let-variables. This decomposition results in \\
$
\left\{~ \fapp{Eval}{\args{\brackets{\var{\Nil}}{:\var{\underline{d_0}}}}{\var{\Prog}}},~ 
\fapp{EvalCall}{\threeargs{\var{\underline{f}}}{\bullet}{\expr{\var{\Prog}}}},~
\fapp{Eval}{\args{\bullet}{\expr{\var{\Prog}}}},~ \bullet~ \right\},~
$ \\
$
\texttt{\hspace{80pt} let}~ \var{e.x} ~\texttt{=}~ \bullet~ ~\texttt{in}~
\left\{~ {
\fapp{Eval}{\args{\brackets{\var{\Nil}}{:\var{\Nil}}}{\expr{\var{\Prog}}}}{},~
\var{e.x}\,\append\, \bullet~
} \right\}
$, where $\expr{\var{e.x}}$ is a fresh parameter.

Hence, considering modulo the arguments, the following holds.
Given a function-call stack element $\expr{\var{f}}$, this $\expr{\var{step}}$ maps the interpreted stack element to this segment of the interpreting function-call stack represented by the first configuration above, 
when this stack segment will be computed then its result is declared as a value of parameter $\expr{\var{e.x}}$ and the last configuration will be unfolded.
Note that 
(1) these two configurations separated with 
the \texttt{let}-construct will be unfolded completely separately one from the other, i.e., the first configuration becomes the input of the unfolding  loop, while the second configuration is postponed for a future unfolding call; 
(2) built-in function append $\append$ is not inserted in the stack at all, since it is treated by the supercompiler as a kind of a special constructor, which properties are known by the supercompiler handling this special constructor on the fly. 
The sequence  $\expr{\var{step}}(K_i),\; \ldots,\; \expr{\var{step}}(K_{j-1})$ between two consecutive applications $\expr{\var{step}}(K_i),\; \expr{\var{step}}(K_j)$ of the first $\expr{\var{Eval}}$ rewriting rule unfolds the big-step  of the interpreter, interpreting the regular step corresponding to the application of a rewriting rule of the $f$ definition interpreted.

Given an expression $\expr{\var{exp}}$ to be interpreted by the interpreter, $\expr{\var{exp}}$ defines the current state of the $\expr{\var{P_0}}$ function-call stack. Let $C$ be the configuration representing this stack state (see Section \ref{subsec:Interpreter}). Let $f(exp_1)$ be the 
application on the top of the stack. Then the current 
$\expr{\var{step}}(K_i)$ corresponding to the application of the first $\expr{\var{Eval}}$ rewriting rule maps $\expr{\var{f}}$ to the stack segment $\expr{\var{Eval_1}}$, $\expr{\var{EvalCall_2}}$, $\expr{\var{Eval_3}}$ of the interpreter, considering modulo their arguments, and this stack segment becomes the leading segment of the interpreting function-call stack. The remainder of the interpreted stack is encoded in the arguments of $\expr{\var{Eval_3}}$, $\expr{\var{Eval_4}}$.

This remark allows us to follow the development of these two stacks in parallel. 
Given the following two parameterized configurations $f(exp_1)$ and $Eval((exp_{env}){\;:\;}\underline{f(}\; exp_1 \;\underline{)}, \var{\Prog} )$ we are going to unfold  these configurations in parallel, step by step. The simpler logic of unfolding 
$f(exp_1)$ will provide hints on the logic of unfolding 
$Eval((exp_{env}){\;:\;}\underline{f(}\; exp_1 \;\underline{)}, \var{\Prog} )$.

Now we consider the set of the configuration pairs that may be generated by the unfold-fold loop and are in the relation {$\turprec$}.

\vspace{-5pt}
\subsection{Internal Properties of the Interpreter Big-Step}\label{subsub:Internal-Properties}

\vspace{-5pt}
In this section we consider several properties of the configurations generated by the unfold-fold loop inside one big-step of the self-interpreter. 
In order to prove indirectly that the program model is safe, we start off by unfolding the following initial configuration
{$
{\fapp{Int}{\args{\expr{\brackets{\expr{\var{Call}}~\expr{\var{Main}}~{\expr{\var{e.d}}}}}}{\expr{\brackets{\expr{\var{Prog}}~\expr{\var{Synapse}}}}}}{}}
$}, 
where the value of {$\expr{\var{e.d}}$} is unknown. Let $\expr{\var{\Syn}}$ stand for $\expr{\brackets{\expr{\var{Prog}}~\expr{\var{Synapse}}}}$.

Consider any configuration 
$C_b$ generated by the unfold-fold loop and initializing a big-step of the interpreter. 
Firstly, we assume that $C_b$ is not generalized and no configuration was generalized by this loop before $C_b$. In such a case, $C_b$ is of the form 
{\small
$ 
\fapp{Eval}{\args{\brackets{\expr{\var{env}}}{:\expr{\var{\underline{arg}}}}}{\expr{\var{\Syn}}}},~ 
\fapp{EvalCall}{\threeargs{\var{\underline{f}}}{\bullet}{\expr{\var{\Syn}}}},~
\fapp{Eval}{\args{\bullet}{\expr{\var{\Syn}}}},~ \ldots
$ }\\
where ${\expr{\var{arg}}}$ stands  
for the formal \emph{syntactic} argument taken from the right-hand side of a rewriting rule where  $\underline{\fapp{f}{\var{arg}}{}}$\,  
originates from, 
$\expr{\var{env}}~{::=}~(~\expr{\var{var}}{\;:\;}\expr{\var{val}}~){\;:\;}\expr{\var{env}} ~|~ \Nil$, $\expr{\var{var}}~{::=}~\underline{\expr{\var{s.n}}} ~|~  \underline{\expr{\var{e.n}}}$, and $\expr{\var{val}}$ stands for a partially known value of variable $\expr{\var{var}}$.
Since application $\underline{\fapp{f}{\var{arg}}{}}$\, is on the top of the stack, argument ${\expr{\var{arg}}}$ includes no function application. 
As a consequence, the leading $\expr{\var{Eval}}$ application has only to look for variables and to call substitution $\expr{\var{Subst}}$ if a variable is encountered. 

Thus, excluding all the transitive configurations encountered before the substitution, we consider the following configuration: 
$
\left\{ 
{\zb}{\zb}\fapp{Subst}{\args{\var{env}}{\brackets{\app{\var{Var}}{\var{\underline{var_{t.n}}}}}{}}}{}, {\zb}\bullet {\zb}\right\}{\hspace{-2pt}}, 
$ 
$
\texttt{let}~
 \var{e.x_1} ~\texttt{=}~ \bullet~ ~\texttt{in}~
 \left\{ 
\left\{ {
\fapp{Eval}{\args{\brackets{\var{env}}{:\expr{\var{\underline{arg_1}}}}}{\expr{\var{\Syn}}}}{}, 
 \bullet 
} \right\}{\hspace{-3pt}},~ \right.
$ \\
$ 
\texttt{\hspace{125pt} let}~ \var{e.x_2} ~\texttt{=}~ \bullet~ ~\texttt{in}~
$ 
$
\left. \left\{~
\fapp{EvalCall}{\threeargs{\var{\underline{f}}}{\var{e.x_1}\,\append\, \var{e.x_2}}{\expr{\var{\Syn}}}},~
\fapp{Eval}{\args{\bullet}{\expr{\var{\Syn}}}},~ \ldots
\right\}\; \right\} 
$ \\
 where $\expr{\var{e.x_1}},\, \expr{\var{e.x_2}}$ are {\zb}fresh {\zb}parameters,
{\zb}${\expr{\var{arg_1}}}$ {\zb}stands {\zb}for a part of {\zb}${\expr{\var{arg}}}$ above to be 
processed, {\zb\zb}and $\underline{\expr{\var{var_{t.n}}}}$ denotes the type and the name of the variable encountered.

We turn now to the first configuration to be unfolded. All configurations unfolded, step by step, from the first configuration are transitive (see Section \ref{subsec:Strategy}) since $\expr{\var{Subst}}$ tests only types and names of the environment variables. Function  $\expr{\var{Subst}}$ is tail-recursive and returns value $\expr{\var{val}}$ asked for. 

We skip transforming these transitive configurations and continue with the next one. \\
$
\left\{~ {
\fapp{Eval}{\args{\brackets{\var{env}}{:\expr{\var{\underline{arg_1}}}}}{\expr{\var{\Syn}}}}{},~
 \bullet~
} \right\},~
$ 
$ 
\texttt{\hspace{0pt}let}~ \var{e.x_2} ~\texttt{=}~ \bullet~ ~\texttt{in}~
$ 
$
\left\{~
\fapp{EvalCall}{\threeargs{\var{\underline{f}}}{\var{val}\,\append\, \var{e.x_2}}{\expr{\var{\Syn}}}},~
\fapp{Eval}{\args{\bullet}{\expr{\var{\Syn}}}},~ \ldots
\right\}
$ \\
By our assumption above, the loop unfolding this first configuration never generates a function application. So the leading configuration proceeds to look for the variables in the same way shown above. 

When $\expr{\var{arg}}$ is entirely processed and all variables occurring in $\expr{\var{arg}}$ are replaced with their partially known values from the environment, then the current configuration looks as follows:\\ 
\text{~\hspace{130pt}}
$
\fapp{EvalCall}{\threeargs{\var{\underline{f}}}{\var{arg_2}}{\expr{\var{\Syn}}}},~
\fapp{Eval}{\args{\bullet}{\expr{\var{\Syn}}}},~ \ldots
$ \\
%
Here expression $\expr{\var{arg_2}}$ is $\expr{\var{\underline{arg}}}\theta$, were $\theta$ is the substitution defined by environment $\expr{\var{env}}$. 
I.\,e., $\expr{\var{arg_2}}$ may include parameters standing for unknown data, while $\expr{\var{\underline{arg}}}$ does not.
Any application of $\expr\var{EvalCall}$
 function is one-step transitive. Recalling $\expr\var{\Syn}$, we turn to the next configuration:\\
\text{~\hspace{40pt}}
$
\fapp{Prog}{\expr{\var{Synapse}}}{},~
\fapp{LookFor}{\args{\expr{\var{\underline{f}}}}{\bullet}},~
\fapp{Matching}{\args{\threeargs{\var{\False}}{\var{\Nil}}{\bullet}}{\expr{\var{arg_2}}}},~
\fapp{Eval}{\args{\bullet}{\expr{\var{\Syn}}}},~ \ldots
$ 

$\fapp{Prog}{\expr{\var{Synapse}}}{}$ returns the source code of the interpreted program  $\expr{\var{Synapse}}$, while the $\expr{\var{LookFor}}$ application returns the definition of the function called by the interpreter, using the known name $\expr{\var{\underline{f}}}$. Skipping the corresponding transitive configurations, we have: \\
\text{~\hspace{60pt}}
$
\fapp{Matching}{\args{\threeargs{\var{\False}}{\var{\Nil}}{\brackets{\brackets{\var{\,\underline{p_1}\,}}{{\;:}\texttt{'='}{\;:\;}\brackets{\,\var{\underline{exp_1}}\,}}}{{\;:\;}\var{\underline{def_{r_1}}}\,}}}{\expr{\var{arg_2}}}},~
\fapp{Eval}{\args{\bullet}{\expr{\var{\Syn}}}},~ \ldots
$ \\
%
Here the third $\expr{\var{Matching}}$ argument is the $\expr{\var{f}}$ definition, where $\expr{\var{p_1}}$, $\expr{\var{exp_1}}$, $\expr{\var{def_{r_1}}}$ stand for the pattern, the right-hand side of the first rewriting rule of the definition, and the rest of this definition, respectively. 
This $\expr{\var{Matching}}$ application transitively initiates matching the parameterized data $\expr{\var{arg_2}}$ against pattern $\expr{\var{\underline{p_1}}}$ and calls another  $\expr{\var{Matching}}$ application.
 This second $\expr{\var{Matching}}$ application is provided with the $\expr{\var{f}}$ definition rest and $\expr{\var{arg_2}}$ for the case this pattern matching will fail. The next configuration is as follows.

\vspace{-15pt}
\begin{gather}\label{conf:Match}
\fapp{Match}{\threeargs{\var{\underline{p_1}\,}}{\var{arg_2}}{\brackets{\var{\Nil}}}},~
\fapp{Matching}{\args{\threeargs{\var{\bullet}}{\var{\underline{exp_1}}}{\var{\underline{def_{r_1}}}\,}}{\expr{\var{arg_2}}}},~
\fapp{Eval}{\args{\bullet}{\expr{\var{\Syn}}}},~ \ldots   
                                            \tag{\checkmark}
\end{gather}

\vspace{-5pt}
\begin{remark}\label{rem:1st-on-transitive}
By now all the configurations generated by the unfolding loop were transitive. The steps
processing syntactic structure of the function application considered might meet constructor applications. These constructor applications are accumulated in the second $\expr\var{EvalCall}$ argument.
The analysis above did not use any particular property of the interpreted program despite the fact that the source code of the interpreted program has been received and processed.
\end{remark}

\vspace{-3pt}
Now we start to deal with function $\expr{\var{Match}}$ playing the main role in our analysis.
In order to unfold the configuration \ref{conf:Match}, 
we have now to use some particular properties of the interpreted program.

Since for any pattern $\expr{\var{p_1}}$ in program Synapse and any $\expr{\var{v}} \in {\cal V}$
$\expr{\var{\mu_v(p_1)}} < 2$ holds, Proposition \ref{prop:Match} below implies that the unfold-fold loop never stops unfolding the configuration \ref{conf:Match} until the $\expr{\var{Match}}$ application on the top of the stack will be completely unfolded to several passive expressions, step by step. 
These expressions will appear on different possible computation paths starting at the configuration above.  Skipping the steps unfolding the tree rooted in this stack-top configuration, we turn to the configurations that appear on the leaves of this tree. Each path starting at the top configuration leads to a configuration of one of the following two forms. These configurations are transitive:\\
%
\text{\hspace{100pt}}
$
\fapp{Matching}{\args{\threeargs{\brackets{\var{env_1}}}{\var{\underline{exp_1}}}{\var{\underline{def_{r_1}}}\,}}{\expr{\var{arg_3}}}},~
\fapp{Eval}{\args{\bullet}{\expr{\var{\Syn}}}},~ \ldots
$   \\
%
\text{\hspace{100pt}}
$
\fapp{Matching}{\args{\threeargs{\var{\False}}{\var{\underline{exp_1}}}{\var{\underline{def_{r_1}}}\,}}{\expr{\var{arg_3}}}},~
\fapp{Eval}{\args{\bullet}{\expr{\var{\Syn}}}},~ \ldots
$

In the first case, the pattern matching did succeed and function $\expr{\var{Matching}}$ replaces the current function application with the right-hand side of the chosen rewriting rule, provided with the constructed environment. The big-step being considered has been finished. 
In order to launch the next big-step, interpreter $\expr{\var{Int}}$ has now to update the top of the interpreting function application stack.

In the second case, the pattern matching fails and function $\expr{\var{Matching}}$ once again calls $\expr{\var{Match}}$, aiming to match the parameterized data against the pattern of the next rewriting rule of function $\expr{\var{f}}$. 
The next configuration is of the form \ref{conf:Match}  above,
 in which the third $\expr{\var{Matching}}$ argument value is decremented with the rewriting rule has been considered.
If this value is empty and cannot be decremented, then, according to the language ${\cal L}$ 
semantics, see Section \ref{subsec:Self-Int-Language}, 
we have the abnormal deadlock state and the interpreter work is interrupted.
Starting off from this configuration, the unfold-fold loop proceeds in the way shown above. 

\vspace{-1pt}
\begin{proposition}\label{prop:Match} 
For any pattern $\expr{\var{p_0}}$ such that for any $\expr{\var{v}} \in {\cal V}$
$\expr{\var{\mu_v}(\var{p_0})} < 2$ and any parameterized passive expression $\expr{\var{d}}$, the unfold-fold loop, starting off from configuration
 $\fapp{Match}{\threeargs{\var{\underline{p_0}\,}}{\var{d}}{\brackets{\var{\Nil}}}}$ 
and controlled by the $\turprec$-strategy, results in a tree program\footnote{I.e., without any function application, except an entry point of this residual program.} such that any non-transitive vertex in the tree is labeled by a configuration of the form
$\fapp{Match}{\threeargs{\var{\underline{p_i}\,}}{\var{d_i}}{\brackets{\var{env_i}}}},~ \ldots~ $.
Given a path in the tree and any two configurations $\var{M_i}$, $\var{M_j}$ of the forms
$\fapp{Match}{\threeargs{\var{\underline{p_i}\,}}{\var{d_i}}{\brackets{\var{env_i}}}},~ \ldots~ $ and
$\fapp{Match}{\threeargs{\var{\underline{p_j}\,}}{\var{d_j}}{\brackets{\var{env_j}}}},~ \ldots~ $,  respectively, belonging to the path, 
such that $\var{M_j}$ is a descendant of $\var{M_i}$,  then 
 $\expr{\var{\underline{p_j}}} \prec \expr{\var{\underline{p_i}}}$ holds.
\end{proposition}
\vspace{-3pt} %
\textbf{Proof} 
If all descendants of configuration  $\fapp{Match}{\threeargs{\var{\underline{p_0}\,}}{\var{d}}{\brackets{\var{\Nil}}}}$
are transitive then the unfold-fold loop results in a tree being a root and this tree satisfies the property required. 
Now  consider non-transitive descendants of $\fapp{Match}{\threeargs{\var{\underline{p_0}\,}}{\var{d}}{\brackets{\var{\Nil}}}}$ that may be generated by the unfold-fold loop before the first generalization or folding action 
happened. The patterns of the $\expr{\var{PutV}}$ rewriting rules never test unknown data, hence any application of $\expr{\var{PutV}}$ is transitive. Since for any $\expr{\var{v}} \in {\cal V}$  relation $\expr{\var{\mu_v}(\var{p_0})} < 2$ holds, 
function $\expr{\var{Eq}}$ will be never applied and the application of $\expr{\var{CheckRepVar}}$ is transitive. 
As a consequence, all the non-transitive descendants are of the forms $\fapp{Match}{\threeargs{\var{\underline{p_i}\,}}{\var{d_i}}{\brackets{\var{env_i}}}}, \ldots~ $.
Only the paths originated by applications of the 2-nd, 3-rd, 4-th $\expr{\var{Match}}$ rewriting rules may contain configurations of such forms. 

Consider a configuration $\fapp{Match}{\threeargs{\var{\underline{p_i}\,}}{\var{d_i}}{\brackets{\var{env_i}}}}, \ldots~ $.
Below $M_i$ denotes such a configuration.
Since $\expr{\var{p_i}}$ is a constant, 
one-step unfolding this configuration by the 2-nd rewriting rule leads to configuration 
$\fapp{PutVar}{\args{\expr{\var{\underline{s.n}}}{\,:\;}\expr{\var{d_s}}}{\brackets{\var{env_i}}{}}},~
\fapp{Match}{\threeargs{\var{\underline{p_{i+1}}\,}}{\var{d_{i+1}}}{\bullet}}, \ldots~ $ such that 
$\expr{\var{\underline{p_{i+1}}}}$ is a proper part of  $\expr{\var{\underline{p_i}}}$, in which at least one constructor is removed. Hence, $\expr{\var{\underline{p_{i+1}}}} \prec \expr{\var{\underline{p_i}}}$ holds.
Since the $\expr{\var{PutVar}}$ application is transitive, a number of unfolding steps lead transitively to 
$\fapp{Match}{\threeargs{\var{\underline{p_{i+1}}\,}}{\var{d_{i+1}}}{\brackets{\var{env_{i+1}}}{}}}, \ldots~ $ such that $\expr{\var{\underline{p_{i+1}}}} \prec \expr{\var{\underline{p_i}}}$.

One-step unfolding the configuration $\fapp{Match}{\threeargs{\var{\underline{p_i}\,}}{\var{d_i}}{\brackets{\var{env_i}}}}, \ldots~ $ by the 3-rd rewriting rule leads to configuration 
$\fapp{Match}{\threeargs{\var{\underline{p_{i+1}}\,}}{\var{d_{i+1}}}{\brackets{\var{env_i}}}},~
\fapp{Match}{\threeargs{\var{\underline{p_{i+2}}\,}}{\var{d_{i+2}}}{\bullet}}, \ldots~ $ such that 
$\expr{\var{\underline{p_{i+1}}}}$ and $\expr{\var{\underline{p_{i+2}}}}$  are proper parts of constant $\expr{\var{\underline{p_i}}}$. Hence, $\expr{\var{\underline{p_{i+1}}}} \prec \expr{\var{\underline{p_i}}}$ and $\expr{\var{\underline{p_{i+2}}}} \prec \expr{\var{\underline{p_i}}}$ hold.

One-step unfolding the configuration $\fapp{Match}{\threeargs{\var{\underline{p_i}\,}}{\var{d_i}}{\brackets{\var{env_i}}}}, \ldots~ $ by the 4-th rewriting rule leads to configuration 
$\fapp{Match}{\threeargs{\var{\underline{p_{i+1}}\,}}{\var{d_{i+1}}}{\brackets{\var{env_i}}}},~ \ldots~ $ such that 
$\expr{\var{\underline{p_{i+1}}}}$ is a proper part of  $\expr{\var{\underline{p_i}}}$, in which at least one constructor is removed. Hence, $\expr{\var{\underline{p_{i+1}}}} \prec \expr{\var{\underline{p_i}}}$ holds.

Now consider any two configurations of the forms
$\fapp{Match}{\threeargs{\var{\underline{p_i}\,}}{\var{d_i}}{\brackets{\var{env_i}}}}, \ldots~ $
and \\
$\fapp{Match}{\threeargs{\var{\underline{p_j}\,}}{\var{d_j}}{\brackets{\var{env_j}}}}, \ldots~ $
such that the second configuration belongs to a path originating from the first one and is encountered before any generalization. Hence, $\expr{\var{\underline{p_i}}}$ and $\expr{\var{\underline{p_j}}}$ are constants.

Given a configuration $C$, the length of $C$, denoted by $ln_c(C)$, is the number of the \emph{upper} function applications in $C$. (See Definition \ref{def:configuration} above.)

If $ln_c(M_j) < ln_c(M_i)$ then $M_i \nottur M_j$ and the Turchin relation  prevents $M_j,\, M_i$ from generalization and $M_j$ from any folding action. (See Section \ref{subsec:Strategy}.)

If  $ln_c(M_j) \ge ln_c(M_i)$ 
and $\var{M_i} \tur \var{M_j}$ 
then we consider the first shortest configuration $M_k$ in the path segment being considered. 
By definition of the {\zb}stack,
$\fapp{Match}{\threeargs{\var{\underline{p_k}\,}}{\var{d_k}}{\bullet\text{{\zb}{\zb}}}}$
is from $M_i$ and {\zb}$\fapp{Match}{\threeargs{\var{\underline{p_j}\,}}{\var{d_j}}{\text{{\hspace{-1pt}}}\brackets{\var{env_j}}\text{{\hspace{-2pt}}}}}$ is a descendent of  $\fapp{Match}{\threeargs{\var{\underline{p_k}\,}}{\var{d_k}}{\brackets{\var{env_k}}}}$.

Since $\fapp{Match}{\threeargs{\var{\underline{p_k}\,}}{\var{d_k}}{\bullet}}$ took a part in computing $M_j$, it is the last function application of the stack prefix defined by the following Turchin relation $M_i \tur M_j$, which holds. 
On the other hand, by the reasoning given above, for any $p_{j_l}$ from the prefix of $M_j$, defined by \emph{this} Turchin relation, $\underline{p_{j_l}} \prec \underline{p_k}$ holds, and, as a consequence,
the following relation $\fapp{Match}{\threeargs{\var{\underline{p_{j_l}}\,}}{\var{d_{j_l}}}{\brackets{\var{env_{j_l}}}}} \prec \fapp{Match}{\threeargs{\var{\underline{p_k}\,}}{\var{d_k}}{\bullet}}$ holds. 
According to 
the $\turprec$-strategy 
this relation prevents $M_j,\, M_i$ from generalization and $M_j$ from any folding action. (See Section \ref{subsec:Strategy}.)
 The proposition has been proven.
\hfill$\Box$

\vspace{-5pt}
\subsection{Dealing with the Interpreter Function-Application Stack}\label{subsub:Int-Stack}

\vspace{-3pt}
Firstly, assume that no generalization happened in the unfold-fold loop up to now.
Given a right-hand side $\expr{\var{exp_0}}$ of a rewriting rule returned by function $\expr{\var{Matching}}$ as described in Section \ref{subsub:Internal-Properties}, then the following configuration has to map, step by step, the segment of the interpreted function-application stack, that is defined by known $\expr{\var{\underline{exp_0}}}$, on the top of the interpreting stack:
%
$
\fapp{Eval}{\args{\brackets{\var{env}}{{\;:\;}\var{\underline{exp_0}}\,}}{\expr{\var{\Syn}}}},~ \ldots
$

According to the call-by-value semantics, function $\expr{\var{Eval}}$ looks for 
the function application, whose right bracket is the leftmost closing bracket, 
in completely known $\expr{\var{\underline{exp_0}}}$. It moves from left to right along $\expr{\var{\underline{exp_0}}}$, substitutes transitively the values of 
the variables encountered, as shown in Section \ref{subsub:Internal-Properties}, pushes the interpreted function application in the interpreting stack, mapping it into an $\expr{\var{EvalCall}}$ application, whenever the interpreted application should be pushed in the interpreted stack. See Section \ref{sub:Meta-Reasoning} for the details. Finally the depth first $\expr{\var{EvalCall}}$ application 
 initiates the next big-step.

Since for any pattern $\expr{\var{p}}$ of program Synapse and any $\expr{\var{v}} \in {\cal V}$ 
$\expr{\var{\mu_v(p)}} < 2$ holds, 
all applications 
of $\expr{\var{Eq}}$, $\expr{\var{ContEq}}$ and $\expr{\var{PutV}}$ 
are transitive. This note together with Proposition \ref{prop:Match} implies 
 Proposition \ref{prop:FirstGener} below.

Let $\expr{\var{p_i}}$ stand for sub-patterns of a pattern of program Synapse, 
$\expr{\var{arg_i}}$ and $\expr{\var{def}}$ stand for partially known parameterized expressions and  several, maybe zero, rewriting rules being a rest of a function definition of Synapse, respectively.
Let $\expr{\var{exp}}$ stand for the right-hand side of a rewriting rule from this definition. 
$\expr{\var{env}}~{::=}~(~\expr{\var{var}}{\;:\;}\expr{\var{val}}~){\;:\;}\expr{\var{env}} ~|~ \Nil$, $\expr{\var{var}}~{::=}~\underline{\expr{\var{s.n}}} ~|~  \underline{\expr{\var{e.n}}}$\,, and $\expr{\var{val}}$ stands for a partially known value of variable $\expr{\var{var}}$.
Let $\expr{\var{Int_0}}$ denote 
{$
{\fapp{Int}{\args{\expr{\brackets{\expr{\var{Call}}~\expr{\var{Main}}~\expr{\var{e.d}}}}}{\expr{\var{\Syn}}}}{}}
$}, 
where the value of {$\expr{\var{e.d}}$} is unknown.

\vspace{-2pt}
\begin{proposition}\label{prop:FirstGener} 
Let the unfold-fold loop be controlled by the $\turprec$-strategy. Let it start off from the initial configuration $\expr{\var{Int_0}}$.
Then the first generalized configuration generated by this loop, if any, will generalize two configurations of the following form  
and any configuration folded, before this generalization, by a previous configuration is of the 
 same form, where $n > 0$,
\vspace{-10pt}
\begin{gather}\label{conf:GenMatching}
\text{\hspace{-8.5pt}}
\fapp{Match}{\threeargs{\var{\underline{p_1}\,}}{\var{arg_1}}{\text{\zb\zb\zb}\brackets{\var{env}}\text{\zb\zb}}}, 
\text{{\small \ldots}},\, 
\fapp{Match}{\threeargs{\var{\underline{p_n}\,}}{\var{arg_n}}{\var{\bullet}}},\, 
\fapp{Matching}{\args{\threeargs{\var{\bullet}}{\var{\underline{exp}}}{\var{\underline{def}}\,}}{\expr{\var{arg_0}}}},\, \tag{\P} 
 \fapp{Eval}{\args{\bullet}{\expr{\var{\Syn}}}}, 
  \text{...}    
       \notag           
\end{gather}

\end{proposition}

\vspace{-2pt}
Now consider any application of the form $\fapp{Match}{\threeargs{\var{\underline{p_1}\,}}{\var{arg_1}}{\brackets{\var{env}}}}$ staying on the stack top. Let $\expr{\var{Match_1}}$ denote this application.
Only the third rewriting rule of $\expr{\var{Match}}$ increases the stack. In this case the next state after the stack \ref{conf:GenMatching} is of the form $\expr{\var{Match_3}},~ \expr{\var{Match_2}}\; \ldots$. 
Then, by Proposition \ref{prop:Match}, along any path originating from $\expr{\var{Match_2}}$ the application $\expr{\var{Match_2}}$ is not replaced until application $\expr{\var{Match_3}}$ will be completely unfolded. 
Hence, for any stack state of the form $f_i,~ \ldots, ~ \expr{\var{Match_2}},~ \ldots$ on such a path, where $f_i$ denotes any function application, the following relation   
$\expr{\var{Match_2}},~ \ldots \nottur f_i,~ \ldots, ~ \expr{\var{Match_2}},~ \ldots$ holds. That proves the following corollary using the notation given above.

\vspace{-5pt}
\begin{corollary}\label{prop:FirstGener2} 
Given a timed application $\fapp{Matching_{t_0}}{\args{\threeargs{\var{\bullet}}{\var{\underline{exp}}}{\var{\underline{def}}\,}}{\expr{\var{arg_0}}}}$, 
 any two timed configurations of the form 
{$
 \ldots,~
\fapp{Matching_{t_0}}{\args{\threeargs{\var{\bullet}}{\var{\underline{exp}}}{\var{\underline{def}}\,}}{\expr{\var{arg_0}}}},~
\fapp{Eval}{\args{\bullet}{\expr{\var{\Syn}}}},~ \ldots   
$}
can neither be generalized nor folded one by the other.
\end{corollary}

\vspace{-5pt}
Since any application {$\fapp{Matching}{\args{\threeargs{\var{\dots}}{\var{\underline{exp}}}{\var{\underline{def}}}}}$} decreases, step by step, the list {$\var{\underline{def}}$} of the rewriting rules, the following corollary holds.

\vspace{-5pt}
\begin{corollary}\label{prop:FirstGener3} 
Given a big-step of {$\var{Int}$}, a timed pair {$\expr{\var{\underline{exp_{t_1}}}},\,\expr{\var{\underline{def_{t_1}}}}$}\footnote{Here we use the notation 
given above and the timed expressions, which are defined in the same way as the timed applications (Section \ref{subsec:Turchin}).}, and a timed application \\ $\fapp{Matching_{\tau_0}}{\args{\threeargs{\var{\bullet}}{\var{\underline{exp_{t_1}}}}{\var{\underline{def_{t_1}}}\,}}{\expr{\var{arg_0}}}}$ inside this big-step, then any two timed configurations of the form \\
{$
 \ldots,~
\fapp{Matching_{\tau_i}}{\args{\threeargs{\var{\bullet}}{\var{\underline{exp_{t_i}}}}{\var{\underline{def_{t_i}}}\,}}{\expr{\var{arg_0}}}},~
\fapp{Eval}{\args{\bullet}{\expr{\var{\Syn}}}},~ \ldots   
$}
can neither be generalized nor folded one by the other.
\end{corollary}

\vspace{-5pt}
Given a function definition $F$ and a rewriting rule $r$ of the definition, let $\expr{\var{exp_{F,r}}}$ stand for the right-hand side of $r$, while $\expr{\var{def_{F,r}}}$ stand for the rest of this definition rules following the rule $r$. 
The following is a simple syntactic property of the Synapse program model given in Section \ref{sec:Specifying-Cache}.
\begin{gather}\label{prop:SyntacticProperty}
\text{\vspace{-15pt}}
 \text{For any~} F_1, F_2 \in \{ \var{Main}, \var{Loop}, \var{Event}, \var{Append}, \var{Test} \} \text{~and for any two \emph{distinct} rewriting rules~} r_1, r_2\\ 
 \text{~of~} F_1, F_2\text{, respectively, } 
(\expr{\var{exp_{F_1,r_1}}}, \expr{\var{def_{F_1,r_1}}}) \neq (\expr{\var{exp_{F_2,r_2}}}, \expr{\var{def_{F_2,r_2}}})
 \text{~holds.} \notag
\end{gather}
 
\vspace{-5pt}
 That together with Corollaries \ref{prop:FirstGener2}, \ref{prop:FirstGener3} imply:

\begin{proposition}\label{prop:NoLocalFirstGener} 
Given two configurations $C_1$ and $C_2$ of the form \ref{conf:GenMatching} to be generalized or folded one by the other. Then (1) $C_1$ and $C_2$ cannot belong to the same big-step of $\var{Int}$; (2) there are two functions $F_1$, $F_2$ of a cache coherence protocol model from the series mentioned in Section \ref{sec:Conclusion} (and specified in the way shown in Section \ref{sec:Specifying-Cache}) 
and rewriting rules $r_1, r_2$ of $F_1, F_2$, respectively, such that 
$(\expr{\var{exp_{F_1,r_1}}}, \expr{\var{def_{F_1,r_1}}}) = (\expr{\var{exp_{F_2,r_2}}}, \expr{\var{def_{F_2,r_2}}})$, where functions $F_1$, $F_2$ and rules $r_1, r_2$ may coincide,  respectively.
\end{proposition}

\begin{remark}\label{rem:2st-on-Synapse-properties}
Proposition \ref{prop:NoLocalFirstGener} depends on Property \ref{prop:SyntacticProperty}. Nevertheless, the restriction imposed by \ref{prop:SyntacticProperty} is very weak and the most of programs written in ${\cal L}$ satisfy \ref{prop:SyntacticProperty}. 
It can easily be overcome by providing the interpreting function $\var{Matching}$ with an additional argument that is the interpreted function name.
The second statement of this proposition is crucial for the expectation of removing the interpretation overheads.
Despite the fact that the reasoning above follows the Synapse program model, it can be applied to any protocol model used in our experiments described in this paper.
\end{remark}

\vspace{-3pt}
\emph{We conclude that any configuration encountered by the loop unfolding the given big-step is neither generalized with another configuration generated in unfolding \emph{this} big-step nor folded by such a configuration.}

\vspace{-10pt}
\subsection{On Generalizing the Interpreter Configurations}\label{sub:On-Generalizing}

The unfold-fold loop processes the paths originating from the initial configuration $\expr{\var{Int_0}}$, following the corresponding interpreted paths. The latter ones are processed according to the order of the rewriting rules in the Synapse function definitions. 
The configuration $\var{Int_0}$ is unfolded according to the $\var{Main}$ function definition of the interpreted program. Application of this function leads transitively to the following function application:  
%
$
\fapp{Loop}{\expr{\brackets{\var{time}}}{\,:\,}\expr{\brackets{\var{Invalid}~\var{I}~\var{is}}}{\,:\,}\expr{\brackets{\var{Dirty}~}}{\,:\,}\expr{\brackets{\var{Valid}~}}{}}
$. 
%
The interpreter has to match this call against the left-hand side of the first rewriting rule of the $\var{Loop}$ definition. The first corresponding pattern is: 
{$\brackets{\texttt{[]}}{\,:\,}{\expr{\brackets{\var{Invalid}~\var{e.is}}}}{\,:\,}{\expr{\brackets{\var{Dirty}~\var{e.ds}}}}{\,:\,}{\expr{\brackets{\var{Valid}~\var{e.vs}}}}$}.

The pattern matching processes the pattern and argument from the left to the right, by means of  function $\var{Match}$. The known part of the tree structure is mapped into the interpreting stack by the third rule of $\var{Match}$. The number of $\var{Match}$ applications in the stack is increased by this rewriting rule.

In the given context of specialization the values of $\var{time}$ and $\var{is}$ are unknown. Hence, the prefix of this stack that is responsible for matching the argument constant structure on the left-hand side of  $\var{time}$ will transitively disappear and the unfolding loop will stop at the configuration of the following form \\
\text{\hspace{17pt}}
$\fapp{Match}{\threeargs{\var{\Nil}}{\var{time}}{\brackets{\var{\Nil}}}},~
\fapp{Match}{\threeargs{\var{\underline{p_2}\,}}{\var{arg_2}}{\var{\bullet}}},~
\fapp{Matching}{\args{\threeargs{\var{\bullet}}{\var{\underline{exp}}}{\var{\underline{def}}\,}}{\expr{\var{arg_0}}}},~ 
\fapp{Eval}{\args{\bullet}{\expr{\var{\Syn}}}},~ \bullet$. \\
Here the leading $\var{Match}$ application meets the unknown data $\var{time}$ and has to match it against $\var{\Nil}$ given in the first argument. The environment in the third argument is empty since no variable was still assigned up to now.
The second $\var{Match}$ application is responsible for matching the suspended part of the input data $\var{arg_2}$ against the rest $\var{\underline{p_2}}$ of the pattern. I.\,e.,  $\var{arg_2}$ equals 
{$\expr{\brackets{\var{Invalid}~\var{I}~\var{is}}}{\,:\,}\expr{\brackets{\var{Dirty}~}}{\,:\,}\expr{\brackets{\var{Valid}~}}$}
and $\var{p_2}$ equals 
{${\expr{\brackets{\var{Invalid}~\var{e.is}}}}{\,:\,}{\expr{\brackets{\var{Dirty}~\var{e.ds}}}}{\,:\,}{\expr{\brackets{\var{Valid}~\var{e.vs}}}}$}.

Now we note that the arguments of all applications of the recursive $\var{Loop}$ and $\var{Append}$ functions have exactly the same constant prefix as considered above. That leads to the following proposition.  

\vspace{-5pt}
\begin{proposition}\label{prop:FirstGener4} 
Let the unfold-fold loop be controlled by the $\turprec$-strategy. Let it start off from the initial configuration $\expr{\var{Int_0}}$.
Then the first generalized configuration generated by this loop, if any, will generalize two configurations of the following forms 
and any configuration folded, before this generalization, by a previous configuration is of 
the same form \\
\text{\hspace{12pt}}
$\fapp{Match}{\threeargs{\var{\Nil}}{\var{arg_1}}{\brackets{\var{\Nil}}}},~
\fapp{Match}{\threeargs{\var{\underline{p_2}\,}}{\var{arg_2}}{\var{\bullet}}},~
\fapp{Matching}{\args{\threeargs{\var{\bullet}}{\var{\underline{exp}}}{\var{\underline{def}}\,}}{\expr{\var{arg_0}}}},~ 
\fapp{Eval}{\args{\bullet}{\expr{\var{\Syn}}}},~ \ldots $ 

\end{proposition}

In the given context of specialization Turchin's relation plays a crucial role in preventing the encountered configurations from generalization (see Proposition \ref{prop:Match}). It never forces decomposing the generalized configurations as might do, in general (see Section \ref{subsec:Turchin}). 
Both the crucial configurations of the unfolding history leading to verification of this program model and some properties of the corresponding generalized configurations may be found in the extended version of this paper published as a preprint \cite{Lis_Nem:Veri-via-Inters-arXiv}. 

\vspace{-5pt}
\section{Conclusion}\label{sec:Conclusion}

\vspace{-5pt}
We have shown that  a combination of the verification via supercompilation method and the first Futamura projection allows us to 
perform verification of the program being interpreted. We discussed  the crucial steps of the supercompilation process involved in the verification of a parameterized cache coherence protocol
used as a case study. 
In the same way   we were able to verify all cache coherence protocols from \cite{Delzanno-Param:00, Delzanno-Futurebus:00}, including 
MSI, MOSI, MESI, MOESI, Illinois University, Berkley RISC, DEC Firefly, IEEE Futurebus+, Xerox PARC Dragon, 
specified in the interpreted language ${\cal L}$. 
Furthermore, we were able to verify the same protocols specified in the language WHILE \cite{Jones:Complex}. The complexity of involved processes is huge and further research is required for their better understanding.   

Our experimental results show that Turchin's supercompilation is able to verify rather complicated program models of non-deterministic parameterized computing systems. The corresponding models used in our experiments are constructed on the base of the well known series of the cache coherence protocols mentioned above. So \emph{they might be new challenges to be verified by program transformation rather than an approach for verifying the protocols themselves}. 
This protocol series was early verified by Delzanno \cite{Delzanno-Param:00} and Esparza et al. \cite{Esparza:1999} in abstract terms of equality and inequality constraints. Using unfold-fold program transformation tools this protocol series was early verified by the supercompiler SCP4 \cite{Lis_Nem:CSR07, Lis_Nem:Programming, Lis_Nem:IJFCS08, Lis_Nem:Protocols07} in terms of a functional programming language, several of these protocols were verified in terms of logic programming
\cite{Roychoudhury:2004,  FioPettProSen:GenVer13}. One may consider the indirect protocol models presented in this paper as a new collection of tests developing the state-of-the-art unfold-fold program transformation.

 The intermediate interpreter considered in this paper specifies the operational semantics of a \emph{Turing-complete} language ${\cal L}$. We have proved several statements 
 on properties of the configurations generated by the unfold-fold cycle in the process specializing {$\var{Int}$} with respect to the cache coherence protocols specified as shown in Section \ref{sec:Specifying-Cache}. 
 The main of them is Proposition \ref{prop:Match}.
 Some of these properties \emph{do not depend on specific protocols} from the considered series, i.\,e., they hold for any protocol specified in the way shown in Section \ref{sec:Specifying-Cache}. 
  That allows us to reason, \emph{in a uniform way}, 
   about a huge number of complicated configurations. 
Note that the programs specifying the protocols include both the function call and constructor application stacks, where the size of the first one is uniformly bounded on 
the value of the input parameter while the second one is not.

As a future work, we would like to address the issue of the description of suitable properties of interpreters to which our uniform reasonings demonstrated in this paper might be applied. 

\paragraph{Acknowledgements:} 
We would like to thank Antonina Nepeivoda and anonymous reviewers for helping to improve this work.

\label{sect:bib}
\bibliographystyle{eptcs}

\bibliography{lisitsa_nemytykh}

\end{document}